\documentclass[10pt]{article}
\usepackage{jheppub}

\usepackage{epsfig}
\usepackage{amssymb}
\usepackage{amsmath}
\usepackage{amsfonts}

\newcommand{\be}{\begin{equation}}
\newcommand{\ee}{\end{equation}}
\newcommand{\bea}{\begin{eqnarray}}
\newcommand{\eea}{\end{eqnarray}}
\newcommand{\bi}{\begin{itemize}}
\newcommand{\ei}{\end{itemize}}
\newcommand{\ben}{\begin{enumerate}}
\newcommand{\een}{\end{enumerate}}

\newcommand{\lp}{\left(}
\newcommand{\rp}{\right)}
\newcommand{\ep}{\epsilon)}

\newcommand{\nn}{\nonumber}

\def\gsim{\mathrel{\rlap{\lower4pt\hbox{\hskip1pt$\sim$}}
    \raise1pt\hbox{$>$}}}         %greater than or approx. symbol
\def\lsim{\mathrel{\rlap{\lower4pt\hbox{\hskip1pt$\sim$}}
    \raise1pt\hbox{$<$}}}         %less than or approx. symbol

\def \ep {\epsilon}
\def \Tr {\text{Tr}}

\def\spaa#1#2{\langle#1 #2 \rangle}
\def\spbb#1#2{\left[ #1 #2 \right]}
\def\spab#1#2#3{\langle #1 #2 #3]}

\begin{document}

\preprint{TTP13-005, SFB/CPP-13-07}

\title{$\mathcal O(\alpha_s^2)$ corrections to fully-differential top quark decays}

\author[1]{Mathias Brucherseifer,}
\author[2]{Fabrizio Caola}
\author[2]{and Kirill Melnikov}

\affiliation[1]{Institut f\"ur Theoretische Teilchenphysik, 
Karlsruhe Institute of Technology (KIT), Germany}

\affiliation[2]{Department of Physics and Astronomy, Johns Hopkins 
University, Baltimore, USA}

\emailAdd{mathias.brucherseifer@kit.edu}
\emailAdd{caola@pha.jhu.edu}
\emailAdd{melnikov@pha.jhu.edu}

\abstract{We describe a calculation of the fully-differential decay rate
of a  top quark to a massless $b$-quark and a lepton pair 
at next-to-next-to-leading 
order in perturbative QCD. Technical details of the calculation 
are discussed and selected results for  kinematic distributions 
are shown.}

\maketitle

\section{Introduction} 
\label{intro}

Top quark physics at the LHC is benefiting from huge samples of events 
that can be used to systematically explore many interesting properties 
of the top quark such as  its mass, its production cross-section 
and the structure of its couplings to other Standard Model particles, 
including  the Higgs boson. Many of such studies  require a good understanding 
of the kinematics of top quark decay products that 
are affected by QCD radiative corrections.

The next-to-leading (NLO)  QCD 
corrections to the top quark total decay width  and many kinematic 
distributions are known since long ago \cite{jk,cz,li}.  The electroweak 
corrections have been computed in Refs.~\cite{ew}.  The fully differential 
computation of corrections to the top quark width at next-to-leading 
order was reported in recent years in 
Refs.~\cite{tram1,Campbell:2005bb,Campbell:2012uf}; 
it was performed in the context of NLO 
QCD calculations for a number of processes in 
top quark physics where radiative corrections
to the production and decay stages are incorporated 
consistently \cite{
tram1,Campbell:2005bb,Campbell:2012uf,
Melnikov:2011qx,
Melnikov:2011ta,
Melnikov:2009dn,Bernreuther:2010ny}.

While the NLO QCD description of many hadron collider processes provides 
quite a reasonable approximation,  for some processes it may be useful 
to improve on that and compute the next-to-next-to-leading order 
(NNLO) QCD corrections. 
Top quark pair production in hadron collisions is one of such processes
and computations of NNLO QCD corrections to $pp \to t \bar t$ are 
well underway \cite{Baernreuther:2012ws,
Czakon:2012zr,
Czakon:2012pz}.
So far these calculations are performed for stable top quarks 
and address the total cross-section but, 
eventually, they will mature enough 
to include also top quark decays in the narrow width approximation.
This approximation, when done consistently, includes radiative 
corrections to production and decay stages and 
 gives  an excellent  description of the process 
$pp \to W^+W^-b \bar b$  at NLO QCD~\cite{Denner:2010jp,Denner:2012yc}, 
as was shown explicitly 
in Ref.~\cite{AlcarazMaestre:2012vp}. There is no doubt that the 
narrow width approximation will work very well also at next-to-next-to-leading 
order, once all relevant corrections 
to the production and decay stages 
are accounted for at a fully-differential level. The goal of this 
paper is to provide a fully-differential NNLO QCD calculation 
for the top quark width. 

The  NNLO QCD radiative corrections to the total width 
of the top quark were computed in a number of papers 
\cite{czm,chet,blok2,blok1} and, more recently, NNLO QCD corrections to 
$W$-helicity fractions were reported in Ref.~\cite{hfrac}. 
However, 
NNLO QCD corrections to other kinematic distributions in top decays 
were unknown.
We note that very recently NNLO QCD radiative 
corrections to top quark decay  at a fully differential 
level were computed 
in Ref.~\cite{Gao:2012ja}, using 
a slicing method inspired by soft-collinear 
effective theory \cite{scet}. 
Our computation
employs a totally different method and, as such, is 
complementary to Ref.~\cite{Gao:2012ja}.

A different  phenomenological motivation for this computation 
comes from a similarity between two processes, 
$t \to b + e^+ + \nu$ and $b \to u + e^- + \bar \nu$. 
This similarity implies that 
if  NNLO QCD radiative corrections are known for an arbitrary invariant 
mass of the lepton pair, i.e. beyond the narrow $W$-width approximation, 
they can be immediately applied to the calculation of NNLO QCD corrections 
to moments of kinematic distributions in the 
$b \to u$ semileptonic transition.  Since the $b \to u$ decay suffers from 
significant background from $b \to c$ transition, 
the inclusive rate for 
$b \to u$ is of little help for improving the precision with which 
the CKM matrix element $|V_{ub}|$ can be extracted from experimental
measurements. 
On the other hand, the knowledge of radiative corrections 
to moments  of kinematic distributions with cuts on 
the electron energy  and hadronic invariant mass 
is useful for constructing a theoretical framework to extract 
$|V_{ub}|$, 
see  e.g. Refs.~\cite{bauer1,bauer2,
ural1,ural2}.  The computational framework that we develop in the current 
paper is immediately applicable to the calculation  of these moments.

The other goal of this paper is to contribute to the development 
of a robust computational scheme that will be applicable to any 
collider process at  NNLO QCD.  
A promising way towards such a general scheme 
was described   by Czakon 
\cite{Czakon:2010td,Czakon:2011ve}
who suggested to combine ideas of sector decomposition 
\cite{binothheinrich1,binothheinrich2,an1}
and  the Frixione-Kunszt-Signer (FKS) 
phase-space partitioning \cite{Frixione:1995ms}. 
This computational scheme was further 
discussed in  Ref.~\cite{Boughezal:2011jf} where it was applied 
to the calculation of  NNLO  QED corrections  to $Z \to e^+e^-$. 
In this paper, we employ  these methods to compute 
NNLO QCD radiative corrections 
to top quark decay.   From the methodological point of view, top quark 
decay  $t \to b + e^+ + \nu$ provides an interesting case where the 
structure of soft and collinear singularities is relatively simple, 
yet a massive  colored  particle in the initial state 
leads to a  computational complexity.

The paper is organized as follows. 
In the next Section, we discuss the general 
set up of the computation.  In Section~\ref{phasespace} we explain 
how the phase-space 
for various subprocesses related to 
top quark decays can be parametrized to enable the extraction 
of infra-red and collinear singularities.
In Section~\ref{limits} we discuss 
the singular limits of various 
amplitudes required for our computation. 
In Section~\ref{correct} we explain some peculiarities that 
arise when we try to keep  all momenta of external 
particles in four dimensions, yet  use  dimensional 
regularization consistently. In Section~\ref{results} 
we describe  the results of the calculation and 
show  some kinematic 
distributions. In Section~\ref{conc} we present our conclusions. 
Finally,  we give explicit 
expressions  for tree amplitudes employed in 
our calculation in Appendix~\ref{tree} 
and show how to compute
one-loop master integrals needed in the constructions 
of one-loop soft limits in Appendix~\ref{oneloopMI}.

\section{The set up}
\label{setup}

As we already mentioned in the Introduction, our goal in this 
paper  is to compute   NNLO QCD corrections 
to the  fully differential  decay rate of the top quark 
$t(p_t) \longrightarrow b(p_b)+ \nu_e(p_5)+ e^+(p_6)
$.
Throughout the paper, 
we work in the top quark rest frame.
The decay to $e^+ \nu$ is facilitated by 
 an emission of a $W$-boson, which can be either on or off 
the mass-shell.
The interaction of the virtual $W$ boson with the neutrino-positron 
pair $W \to \nu (p_5) + e^+(p_6)$ is described by the vertex 
\be
\frac{i g_W}{\sqrt2} \langle  5 | {\gamma^\mu} | 6 ],
\ee
where we use the bra and ket notation for 
massless spinors with definite helicities. We employ the standard notation 
$u_+(p_i) = |p_i,+ \rangle = |i \rangle $, $u_-(p_i) = |p_i,- \rangle = |i ]$ 
and note that  further discussion of spinor-helicity 
methods can be found in the  review  Ref.~\cite{Dixon:1996wi}.

We are interested in the NNLO QCD corrections 
to semileptonic top quark decay. To compute these radiative corrections, 
we require the following ingredients
\begin{itemize} 
\item  two-loop virtual corrections to $t\to b +\nu+ e^+$;
\item  one-loop virtual corrections to $t \to b+ g+ \nu +e^+$; 
\item  tree-level 
double-real emission contributions  to the decay rate, $t \to b +g +g +\nu+ e^+$ and  
$t \to b + q \bar q +\nu+ e^+$.   
\end{itemize} 

We note that most  of the required contributions 
are  available  in the literature. Indeed, 
two-loop QCD virtual corrections to   $t \to Wb$  can be obtained  
from an analytic  computation of similar corrections to $b \to W^* u$,  
reported recently by a number of authors \cite{bon,bell,astr,ben,Huber:2009se}.
One-loop  amplitudes for the $t \to b+ g+ \nu +e^+$ process  can 
be extracted from Ref.~\cite{Campbell:2005bb}.  Double-real 
emission amplitudes for gluons and quarks are not known in a compact analytic 
form but can be easily calculated using the spinor-helicity formalism.

However, similar to other NNLO computations, 
the availability of required  amplitudes  
does not make it obvious how to arrive at a physical answer. 
This  happens because of 
 infra-red and collinear singularities that need to 
be extracted 
from individual 
contributions before integrating over phase-spaces of final-state 
particles.  Until recently, the two methods\footnote{Other approaches to NNLO calculations 
are discussed in~\cite{Catani:2007vq,Weinzierl:2003fx,Somogyi:2010ud}. Some phenomenological applications can be found in~\cite{NNLOpheno}.} that were 
used to extract   singularities 
in  practical  NNLO computations  were sector decomposition 
\cite{binothheinrich1,binothheinrich2,an1}
and antenna subtraction \cite{GehrmannDeRidder:2005cm,Daleo:2006xa}.
Both of these methods 
were developed further in recent years 
\cite{Anastasiou:2010pw,GehrmannDeRidder:2012ja}
with the goal to make them applicable to 
more complex processes than previously considered.

The  method that we use in this paper was suggested 
by Czakon  
\cite{Czakon:2010td,Czakon:2011ve} (see also \cite{Boughezal:2011jf}), 
who pointed out that a combination of sector decomposition 
\cite{binothheinrich1,binothheinrich2,an1}
and  the FKS 
phase-space partitioning \cite{Frixione:1995ms} allows 
to resolve overlapping soft and collinear singularities in a universal way. 
We note that from the technical point of view, 
the case of top quark decay 
is particularly simple since no complicated 
phase-space partitioning is, in fact,  required. 
Indeed, singularities in matrix elements occur when emitted 
gluons are either soft or collinear to other massless partons. 
Since there  are only three strongly interacting massless 
particles in   double-real emission  amplitudes for top decays, 
there is just one  triple-collinear limit and no need for angular 
partitioning arises.  
 The soft singularities  are disentangled by ordering 
gluons  with respect to  their energies and by 
removing the corresponding 
$1/2!$ factor from the phase-space.   
Note that this argument needs to be modified 
for sub-processes with 
two quarks in the final state and  we explain how to do it 
in what follows. 

We now remind the reader about the general approach to fully differential 
NNLO QCD  computations based on sector decomposition and phase-space 
partitioning 
\cite{Czakon:2010td,Czakon:2011ve} (see also \cite{Boughezal:2011jf}). 
The main  idea is to express the phase-space for the final state particles 
through suitable unit-hypercube variables 
that make the singular behavior of scattering 
amplitudes manifest.  Once this is accomplished, it is easy to 
construct  subtraction terms for  real emission 
processes by using expansion in plus-distributions.  
To be specific, we consider a double-real emission contribution 
to the differential decay 
rate of the top quark 
\be
\Gamma^{(RR)}_t = \int {\rm d Lips}~|{\cal M}(\{p\})|^2 J(\{p\}),
\label{eq0}
\ee
where  ${\rm dLips}$ is a phase-space for $t \to b + g+g + e^+ + \nu$,
${\cal M}(\{p\})$ is the decay amplitude and $J(\{ p \})$ is a 
measurement function 
that depends on momenta of final state particles $\{ p \}$.   We assume 
that a  map of  the phase-space onto  a unit hypercube is found; 
variables that parametrize 
the  map are  denoted as $\{ x \}$. 
As we will show below, it is possible to re-write 
Eq.(\ref{eq0}) as 
\be
\Gamma_t^{(RR)} = 
\sum \limits_{i=1}^{N_S} \Gamma_t^{(RR,i)},
\;\;\;
\Gamma_t^{(RR,i)} =  
\prod \limits_{j=1}^{N_{\rm ps}}  
\int \limits_{0}^{1}
\frac{{\rm d} x_j }{x_j^{1+a_{j}^{(i)} \ep} } F_i(\ep,\{x \}) 
\label{eq2}
\ee
where $N_S$ is the number of sectors, 
$\ep = (4-d)/2$ is the parameter of dimensional regularization, 
$N_{\rm ps} $ is the number of independent 
variables required to parametrize the phase-space, 
and 
the function $F$ is the product of the scattering amplitude squared, 
sector-dependent powers of singular variables and the phase-space weight
${\rm Ps}_w(\{x\})$ that may include the measurement function
\be
F_i(\{x\}) = x_1^{b_1} x_2^{b_2} .. x_{N_{ps}}^{b_{N_{ps}}}|{\cal M}(\{p\})|^2
~{\rm Ps}_w(\{x\}).
\ee   
The important  property of these functions  $F_i$ 
 is that limits exist at  each potentially 
singular point $x_j = 0$,
\be 
\lim_{x_j =0} F_i(..,x_j,..) = F_i(..,0,..) \ne \infty.
\label{eq_lim}
\ee 
The existence of limits Eq.(\ref{eq_lim})
allows us to re-write Eq.(\ref{eq2}) in an explicitly integrable 
form by employing the decomposition in plus-distributions 
\be
\frac{1}{x^{1+a \epsilon} } = -\frac{1}{a\epsilon} \delta(x) 
+ \sum \limits_{n=0}^{\infty} \frac{(- \epsilon a)^n}{n!} \left [\frac{\ln^n(x)}{x} 
\right ]_+.  
\ee
Substituting this expansion into Eq.(\ref{eq2}) and collecting 
terms of the same order in $\epsilon$, we obtain 
explicitly  integrable terms that can be divided into two broad 
categories,
\be
\int \limits_{0}^{1} {\rm d} x_i \; \frac{\ln^n(x_i) \left ( F(..,x_i,..) 
 -F(..,0,.. ) \right ) }{x_i}
\;\;\;\;\;\;
{\rm and} 
\;\;\;\;
\int \limits_{0}^{1} {\rm d} x_i \delta(x_i) F(..,0,..).
\label{eq5}
\ee
It is easy to recognize that the first 
integral  in the above equation 
contains  a ``subtraction term'' and the second  integral 
 defines  an ``integrated subtraction 
term'', 
if the language familiar from 
NLO QCD computations 
is borrowed.  It is remarkable that, in the described 
framework,   
the subtraction terms are obtained automatically, once a proper parametrization 
of the phase-space is found and the expansion in plus-distributions 
is performed. 
We also note that any subtraction term  depends on  the  function 
$F(x_1,..,x_{N_{ps}})$ at such values of $x_i$  for which at least one of the 
singular variables vanishes. 
 These kinematic points correspond 
to {\it universal soft and collinear limits} 
of scattering amplitudes so that $F(..,x_i=0,..)$ 
can be calculated in a  process-independent way. 
In what follows, we discuss essential elements of this construction such 
as the phase-space parametrization and singular limits of amplitudes. 

Before proceeding to  that discussion, we explain  some subtleties 
that arise when {\it one-loop} corrections 
to the process $t \to b + g + e^+ + \nu$  are analyzed within this framework. 
In this case, 
the singular part of the phase-space is parametrized in terms 
of two variables, 
$x_1$ and $x_2$, that describe  the energy  of the gluon $g$ 
and the  relative angle between the gluon and the $b$-quark, respectively.
The contribution to the top quark decay rate is 
given by the integral 
\be
\int \limits_{0}^{1}  \frac{{\rm d} x_1 }{x_1^{1+2 \ep}}
\frac{{\rm d} x_2 }{x_2^{1+\ep}} \; [x_1^2 x_2] {2 \rm Re}
\left [ {\cal M}_{t \to b g e^+ \nu}^{\rm 1-loop} 
{\cal M}_{t \to b g e^+ \nu}^{(0),*} \right ]...,
\label{eq_rv}
\ee
where ellipses stand for other non-singular phase-space factors. 
While the integrand in  Eq.(\ref{eq_rv}) suggests that singularities 
can be isolated by constructing an expansion in plus-distributions, it is 
in fact not possible to do that  
right away. The reason is that the one-loop amplitude 
${\cal M}_{t \to b g e^+ \nu}^{\rm 1-loop} $ has {\it branch cuts} 
in the limits $x_1 \to 0$ and $x_2 \to 0$ and we must account for 
them  when constructing the
 expansion of the integrand in plus-distributions. To accomplish that, 
we  write 
\be
[ x_1^2 x_2 ] \left [ {\cal M}_{t \to b g e^+ \nu}^{\rm 1-loop} 
{\cal M}_{t \to b g e^+ \nu}^{(0),*} \right ] 
\sim F_1(x_1,x_2,..) + (x_1^2 x_2)^{-\ep} F_2(x_1,x_2,..),
\label{eq_rv1}
\ee
where the two functions $F_{1,2}$ are {\it free}  from branch cuts 
in $x_{1,2} \to 0$ limits. 

The calculation of real-virtual corrections now proceeds as follows. 
We insert Eq.(\ref{eq_rv1}) into Eq.(\ref{eq_rv}), combine non-integer 
powers of $x_{1,2}$ and expand in plus-distributions. Upon doing so, we 
find a variety of terms that are  similar to what we showed in
Eq.(\ref{eq5}). However, in this case the integrand 
is written in terms of 
$F_1(x_1,x_2)$ and $F_{2}(x_1,x_2)$ and we need a prescription that 
will allow us to compute these functions from scattering amplitudes.
To understand how this is done, 
note that $F_{1,2}(\{x\})$ may appear in  the integrand 
with one or both of its arguments set to zero,  which corresponds to 
soft or collinear kinematics of the final state 
gluon.
To compute $F_{1,2}$ in those cases, we need to investigate 
$\left [ {\cal M}_{t \to b g e^+ \nu}^{\rm 1-loop} 
{\cal M}_{t \to b g e^+ \nu}^{(0),*} \right ] $ in soft and collinear   
limits. 
Consider first the soft limit, $x_1 \to 0$.  In this case, the 
interference of the one-loop and tree matrix elements reads
\be
\begin{split} 
\lim_{x_1 \to 0}^{} 
2 {\rm Re}\left [ {\cal M}_{t \to b g e^+ \nu}^{\rm 1-loop} 
{\cal M}_{t \to b g e^+ \nu}^{(0),*} \right ]  
& =
g_s^2 C_F S_{tb} \lp
2{\rm Re}\left [ {\cal M}_{t \to b e^+ \nu}^{\rm 1-loop} 
{\cal M}_{t \to b e^+ \nu}^{(0),*} \right ]  \rp \\
& +
g_s^2 C_A S^{(1)}_{tb}
\left [ {\cal M}_{t \to b  e^+ \nu}^{(0)} 
{\cal M}_{t \to b  e^+ \nu}^{(0),*} \right ],  
\label{eq_rv2}
\end{split} 
\ee
where $S_{tb}$ and $S_{tb}^{(1)}$ are tree and one-loop eikonal 
factors, respectively. We note that 
in  the right hand side of Eq.(\ref{eq_rv2}), 
the   dependence of the one-loop amplitude 
on the momentum of the soft gluon is entirely 
contained in the eikonal  factors. 
To identify $F_{1,2}$ at vanishing value of $x_1$, it is important to realize that 
the leading-order eikonal factor $S_{tb}$ is free from branch cuts 
while the one-loop eikonal factor  $S_{tb}^{(1)}$
may (and in fact does) contain them.  We conclude that, in the 
soft limit $x_1 \to 0$, the first term 
on the right hand  side in Eq.(\ref{eq_rv2}) matches to 
$F_1(0,x_2)$ and the second term to  $F_2(x_1=0,x_2)(x_1^2 x_2)^{-\ep}$.

A similar situation occurs in the collinear limit $x_2 \to 0$. In this case 
\be
\begin{split} 
\lim_{x_2 \to 0}^{} 
\left [ {\cal M}_{t \to b g e^+ \nu}^{\rm 1-loop} 
{\cal M}_{t \to b g e^+ \nu}^{(0),*} \right ]  
%& 
= \frac{g_s^2}{p_b\cdot p_g}\lp
P^{(0)} _{b \to bg}
\left [ {\cal M}_{t \to {\tilde b} e^+ \nu}^{\rm 1-loop} 
{\cal M}_{t \to {\tilde b} e^+ \nu}^{(0),*} \right ]  
%& 
+
P_{b \to bg}^{(1)}
\left [ {\cal M}_{t \to {\tilde b}  e^+ \nu}^{(0)} 
{\cal M}_{t \to {\tilde b}  e^+ \nu}^{(0),*} \right ]\rp,
\label{eq_rv3}
\end{split} 
\ee
where $P_{b \to bg}^{(0,1)}$ are the tree- and one-loop splitting
functions; these functions can be derived 
from  collinear limits of scattering amplitudes 
described in   Ref.~\cite{Kosower:1999rx}. Similar to the soft limit, 
the tree-level splitting function $P_{b \to bg}^{(0)}$ does not contain any
branch cut in  $x_2 \to 0$ limit, while the one-loop splitting function 
does. Hence, the first term on the right-hand side 
in Eq.(\ref{eq_rv3}) is identified with $F_1(x_1,x_2=0)$ 
and the second one with $(x_1^2 x_2)^{-\ep} F_2(x_1,x_2 = 0)$.

If  $x_1$ and $x_2$ are {\it both} non-vanishing,
we can re-write  combinations of $F_1(x_1,x_2)$ and $F_2(x_1,x_2)$
that appear in the integrand for the rate,  through 
the  one-loop matrix  element interfered 
with the  tree-level amplitude, using  Eq.(\ref{eq_rv1}). 
The one-loop amplitude for the $t \to b + g + e^+ + \nu $
that we require,  can be extracted from 
Ref.~\cite{Campbell:2005bb}, where $Wt$ production in 
$bg$ fusion is discussed.\footnote{
We use  the {\sf Fortran77}  implementation of this amplitude 
in the MCFM program~\cite{mcfm}.  We slightly extend it to allow the 
use of quadruple precision in {\sf Fortran90}.} 
For our purpose, the relevant amplitudes 
need to be crossed to describe the $t \to b+g+W^+$ channel. 
Finally, 
the evaluation of the one-loop amplitude for $t \to b+g + e^+ + \nu$ process 
requires  master integrals; they can  be  computed using the program 
QCDLoops \cite{Ellis:2007qk}. We found, 
however, that results of this program become 
unreliable in extreme kinematic  limits; for example when the momentum of the 
gluon in the final state becomes very small $E_g \lsim 10^{-8} m_t$. 
To overcome this difficulty, we independently 
implemented all  master  integrals
relevant for the computation of  the one-loop 
amplitude for $t \to b+g + e^+ + \nu$
using analytic formulas provided 
in Refs.~\cite{Ellis:2007qk,denner}.
Our implementation allows us to 
employ quadruple precision and to reach 
very  small gluon  energies $E_g \sim 10^{-12} m_t$ and 
very small opening angles between the bottom quark and the gluon.

The importance of reaching small energies and angles follows from 
Eq.(\ref{eq5}) which shows that integration over hypercube variables $\{x \}$ 
should, in principle,  be performed on the interval $x_i \in [0,1]$.
This implies that 
full matrix elements that contribute to the function 
$F(x)$ must be evaluated at vanishingly small values of $x$.  
It is, of course, impossible to compute them at $x=0$ and we need 
to introduce additional regularization. 
Following Refs.~\cite{Czakon:2010td,Czakon:2011ve}
we provide this 
regularization by discarding such points in $\{x\}$-space  
where  the product of singular variables is smaller 
than a user-defined parameter $\delta$. For example, 
for double-real emission 
 corrections where four variables are singular, we require 
$x_1 x_2 x_3 x_4 > \delta$, while for the single-emission contributions we require 
$x_1 x_2 > \delta$.  Values of $\delta$ that we use range from 
$10^{-7}$ to $10^{-12}$,  where extremely small values are employed to test 
the independence of final results on the  cut-off parameter and the
cancellation of infra-red divergences. For phenomenological 
calculations we find that values of $ \delta \sim 10^{-7}-10^{-9}$ are 
small enough to provide reasonable results for the second order corrections.

Finally, we note that Eq.(\ref{eq5}) allows us to compute arbitrary 
kinematic distributions in a single run of the code.  
We only need to determine the kinematics of the event -- which 
is uniquely defined for a given set of $\{x\}$-variables -- 
and assign the relevant weight  to a particular  bin in a histogram 
that we want to compute.  We do this each time the function $F(\{x\})$ 
is  calculated. 
Such  implementation is standard for how subtraction procedure 
is employed  in parton level Monte-Carlo event generators both at 
next-to-leading order (see e.g. Ref.~\cite{mcfm} )  
and beyond \cite{Gavin:2010az}.

\section{Phase-spaces for various sub-processes in top decays}
\label{phasespace}

In this Section, we discuss the phase-space parametrization for the various 
sub-processes related to top quark decays.  We start with the leading order 
phase-space and then proceed to phase-spaces with one and two additional 
gluons in the final state. We also address the parametrization of the 
phase-space for $t \to b + q \bar q + e^+ + \nu$ since it differs from 
the two-gluon case in important details. 

The {\it leading order process} 
 that we deal with is $t \to b + W^* ( e^+ + \nu)$, 
where the $W$-boson can be off the mass shell.  While the $W$ off-shellness 
is not an important effect for top quark decays, our  reasons  for doing 
the calculation this way is explained in the Introduction. 
To account for the $W$ off-shellness in 
a systematic way, it is convenient to include the Breit-Wigner propagator 
from the scattering amplitude into the definition of the phase-space. 
We write 
\be
{\rm d} \Pi_{be^+\nu}  
=  \frac{ {\rm d} p_W^2 }{2\pi}
\frac{ {\rm Lips}(t\to bW^*)  {\rm dLips}(W^* \to \nu l^+)
}{(p_W^2-m_W^2)^2+m_W^2\Gamma_W^2},
\ee
where ${\rm  Lips}$ stands for  Lorentz-invariant phase-spaces for 
indicated final states,
$p_W^2$ is the invariant mass of the lepton pair and $\Gamma_W$ is the 
$W$-decay width.  We remove the 
Breit-Wigner distribution by changing variables 
\be
p_W^2 = m_W^2 + m_W\Gamma_W  \tan \xi,
\ee
so that 
\be
\frac{1}{2\pi}\frac{{\rm d} p_W^2}{(p_W^2-m_W^2)^2+m_W^2\Gamma_W^2} = 
\frac{{\rm d} \xi}{2\pi m_W\Gamma_W}. 
\ee
The kinematics of the process implies a restriction on the invariant mass of 
the lepton pair,  \mbox{$0<p_W^2<m_t^2$}. To account for  it, we re-map 
$\xi$ onto the $(0,1)$ interval by defining 
$\xi = \xi_{\rm min} + (\xi_{\rm max}-\xi_{\rm min}) x_7$. We obtain  
\be
\frac{1}{2\pi}\frac{{\rm d} p_W^2}{(p_W^2-m_W^2)^2+m_W^2\Gamma_W^2}
= N_{\rm BW}\; {\rm d} x_7,
\label{eq1}
\ee
where 
\be
N_{\rm BW} = 
\frac{\xi_{\rm max}-\xi_{\rm min}}{2\pi m_W\Gamma_W},
\;\;\;
\xi_{\rm min} = -\arctan \frac{m_W}{\Gamma_W},
\;\;\;
\xi_{\rm max} = \arctan \left [ \frac{m_t^2-m_W^2}{m_W \Gamma_W} \right ].
\ee
After generating the invariant
mass of the lepton pair using Eq.(\ref{eq1}) , 
we obtain  lepton and neutrino momenta in the $W^*$  
rest frame in the standard way
\be
p_{\nu,l} = \frac{\sqrt{p_W^2}}{2}(1,\pm \vec n_l),
\ee
where $\vec n_l = ( \sin\theta_5\cos\varphi_5,\sin\theta_5\sin\varphi_5,
\cos\theta_5 ) $.
Redefining angular variables, 
\be
\cos\theta_5 = 1-2 x_8,~~~~~~~\varphi_5 = 2\pi x_9,
\ee
we re-write the dilepton phase-space as 
\be
{\rm dLips}(W^* \to \nu l^+) = \frac{{\rm d}x_8 {\rm d} x_9}{8\pi}.
\ee
We note that we treat  the lepton-neutrino phase-space 
as  four-dimensional, even in higher orders of QCD perturbation theory  
where infra-red and collinear divergences require  analytic continuation 
of the space-time dimensionality.  
It is intuitively clear that it should be possible to do so because
lepton momenta are infra-red safe observables. 
However, some subtleties arise when the computation is organized 
this way. We discuss them in Section~\ref{correct}.

Finally, we describe the parametrization  of the $t\to b W^*$ 
phase-space.
We work in the rest frame of the decaying top quark and define the $z$-axis 
by aligning it with the momentum of the $b$-quark.  The momenta read
\be
p_t = m_t(1,0,0,0),\;\;\;\;\;p_b = E_b(1,\vec n_b ),
\;\;\; \vec n_b = (0,0,1).
\ee
The energy of the $b$-quark is given by 
\be
E_b = \frac{m_t^2-p_W^2}{2m_t}. 
\ee
It is easy to find the leading order phase-space 
for $t \to b W^*$ decay. It reads 
\be
{\rm Lips}(t\to b W) = 
\frac{1}{8\pi}\lp 1-\frac{p_W^2}{m_t^2}\rp \Upsilon(\ep) 
= \frac{E_{\rm max}}{4\pi m_t}   \Upsilon(\ep),
\ee
where $E_{\rm max} = E_b$ is the maximal energy of the $b$-quark for 
fixed 
invariant mass of the lepton pair  and  
\be
\Upsilon(\ep) 
= \frac{\Gamma(1-\ep)}{\Gamma(2-2\ep)} 
\left ( \frac{E_{\rm max}^2}{\pi} \right )^{-\ep} = 1 + \mathcal O(\ep).
\ee
We can then write the leading order  phase-space 
for $t \to b + e^+ + \nu$  as
\be
\begin{split}
& {\rm d} \Pi_{b e^+ \nu} = \Upsilon(\ep)N_{\rm BW}
\frac{E_{\rm max}}{32\pi^2 m_t} {\rm d}x_7 {\rm d}x_8 {\rm d}x_9 
%\\
%& 
=
\frac{\Upsilon(\ep)N_{\rm BW}}{64\pi^2} \lp 1-\frac{p_W^2}{m_t^2}\rp {\rm d}x_7 {\rm d}x_8 {\rm d}x_9.
\end{split}
\ee
The $\ep$-dependent part of the leading order phase-space  
$\Upsilon(\ep)$ can be neglected provided that {\it exactly} 
the same 
quantity is  identified and removed from phase-spaces in higher orders 
of perturbation theory.  In our calculation, we set  
$\Upsilon(\ep)=1$ consistently everywhere. 

%\subsection{The phase-space for $t \to b + g(p_3)  + e^+ + \nu$}
We now turn to the phase-space which is relevant 
for the description of the single  emission process 
$ t \to b + g(p_3) + e^+ + \nu$.
The treatment of the $W^*$ 
decay phase-space 
and the Breit-Wigner 
factor is identical to the leading order  case; the new element 
is the phase-space for the process $t \to W^* + b + g$,  which 
we refer to as  ${\rm dLips}(t\to bgW^*)$.  This phase-space 
is split into ``regular'' and ``singular'' parts
\be
{\rm dLips}(t\to bgW^*) = {\rm dLips}(Q \to bW^*) \times [{\rm d}p_3],
\ee
where $Q = p_t - p_3$, 
$[{\rm d}p_3] = {\rm d}^{D-1} p_3/[(2\pi)^{D-1} 2 E_3]$ and 
\be
{\rm dLips}(Q \to bW^*) = 
(2 \pi)^{D} \delta^{(D)}(Q - p_b - p_W) 
[{\rm d}  p_b] [{\rm d} p_W].
\ee
The  momenta that enter the ``regular'' phase-space 
read 
\be
p_t = m_t(1,0,0,0),
\;\;\;
p_b = E_b(1,\vec n_b),
\;\;\;
E_b = \frac{Q^2-p_W^2}{2(Q_0- \vec Q \cdot \vec n_b)}.
\ee
We find 
\be
{\rm dLips}(Q \to bW^*) = 
\frac{\Upsilon(\ep) }{4\pi} \frac{E_b}{Q_0- \vec Q \cdot \vec n_b}\lp 
\frac{E_b}{E_{\rm max}}\rp^{-2\ep}.
\label{eqlpq}
\ee
The gluon momentum is parametrized 
as 
\be
p_3 = E_3(1,\sin\theta_3\cos\varphi_3,\sin\theta_3\sin\varphi_3,\cos\theta_3),
\ee
where gluon energy 
and emission angles are computed from unit hypercube variables 
\be
E_3 = x_1 E_{\rm max},
\;\;\;
\cos\theta_3 = 1-2 x_2,\;\;\;
\varphi_3 = 2\pi x_3.
\ee
Expressed through these variables, the singular phase-space 
 becomes 
\be
[{\rm d}p_3] = 
\frac{\Gamma(1-\ep) {\rm PS}^{-\ep}
}{8 \pi^2 (4\pi)^{-\ep} \Gamma(1-2\ep)}  2 E_{\rm max}^2 
\frac{{\rm d}x_1 {\rm d}x_2 {\rm d}x_3 }{x_1^{1+2\ep} x_2^{1+\ep}}
\left [ x_1^2 x_2 \right ],
\label{eq3.20}
\ee
where ${\rm PS} = 16 E_{ \rm max}^2 (1-x_2)\sin^2\varphi_3$. We note that 
the term $[ x_1^2 x_2 ]$  in Eq.(\ref{eq3.20}) 
gets combined with the matrix element squared  
$|{\cal M}_{t  \to b + e^+ + \nu + g}|^2$ to render the function $F(x)$, that 
we discussed in the previous Section.  
It is this factor  from the phase-space 
that makes $F(x)$ finite both in the soft $x_1 \to 0$ 
and in the collinear $x_2 \to 0$ limits.  
We find it convenient to  factor out 
a term $\Gamma(1+\ep)/[ 8 \pi^2 (4\pi)^{-\ep}] $ 
per order in perturbation theory and to write the 
final result for the differential top quark 
width as a series in $\alpha_s/(2\pi)$.
We therefore write the phase-space for  the process 
$t \to b + g(p_3) + e^+ + \nu$ as
\be
\begin{split}
& {\rm d} \Pi_{bge^+\nu} = 
\Upsilon(\ep) \frac{\Gamma(1+\ep)}{8 \pi^2 (4\pi)^{-\ep}}  
N_{\rm BW} 
\lp\frac{\Gamma(1-\ep)}{\Gamma(1+\ep)\Gamma(1-2\ep)}\rp  \\
&\times \frac{E_{\rm max}^2}{16\pi^2} \frac{E_b}{Q_0- \vec Q \cdot \vec n_b } 
\lp \frac{E_b}{E_{max}}\rp^{-2\ep}
{\rm PS}^{-\ep}   
\frac{{\rm d}x_1}{x_1^{1+2\ep}} \frac{{\rm d}x_2}{x_2^{1+\ep}} 
\; {\rm d}x_3 {\rm d}x_7 {\rm d}x_8 {\rm d}x_9 \; [ x_1^2 x_2 ],
\end{split} 
\label{eq_g}
\ee
where $Q = p_t - p_3$.

Finally, we discuss the phase-space for the {\it double-real emission process}
$t \to b + g(p_3) + g(p_4) + e^+ + \nu$. Similar to the 
single-gluon case that we just discussed, the  phase-space is split 
into a regular  and a singular part. The regular phase-space 
is identical to the single gluon emission case
Eq.(\ref{eqlpq}), up to a change in the 
definition of the momentum $Q = p_t - p_{3} - p_{4}$.  The singular 
phase-space is given by the product of single-particle phase-spaces 
of the two gluons, $g(p_3)$ and $g(p_4)$.  We choose the momenta to be 
\be
\begin{split} 
& p_t = m_t(1,0,0,0),
\;\;\;
p_b = E_b(1,\vec n_b),
\;\;\;
p_3 = E_3(1,\vec n_3 ),\;\;\; 
p_4 = E_4 ( 1, \vec n_4),
\\
& \vec n_3 = (\sin\theta_3\cos\varphi_3,\sin\theta_3\sin\varphi_3,\cos\theta_3),
\;\;\;
\vec n_4 = 
(\sin\theta_4\cos\varphi_{34},
\sin\theta_4\sin\varphi_{34},\cos\theta_4),
\end{split} 
\ee
with $\varphi_{34} = \varphi_3+\varphi_4$. 
We compute  the azimuthal angle $\varphi_3$ using
$\varphi_3 = 2\pi x_6$. We note that the number of particles that 
participate in the ``QCD part'' of the  
process is sufficiently small, so that 
we can embed momenta of  top and bottom quarks and of the 
two gluons into a four-dimensional space-time. By momentum 
conservation,  the momentum of the $W$-boson is also  four-dimensional, 
but this does not imply that momenta of a lepton and 
a neutrino can be chosen to be  four-dimensional as well. We will argue
in Section~\ref{correct} 
that it is in fact possible to do that provided that 
 {\it a spin-correlation part of certain splitting  functions is modified}. 

The correct choice of phase-space variables is crucial for extracting 
singularities from products of  matrix elements squared 
 and phase-space 
factors.  For the two-gluon case, we must consider the following singular 
limits: {\it i}) 
one or both gluons are soft; {\it ii}) one or both gluons are collinear to a 
$b$-quark and {\it iii})  
two gluons are collinear to each other. The parametrization 
of phase-space that leads to factorization of these singularities was 
given in Refs.~\cite{Czakon:2010td,Czakon:2011ve} and we now discuss 
it in the context of  top quark decay.  We note that 
by ordering two gluons in energy, 
we remove the $2!$ symmetry factor from the phase-space.  
We write the singular phase-space as 
\be
\begin{split} 
& [{\rm d} p_3] [{\rm d}p_4] \theta(E_{3} - E_{4}) 
= 
\frac{{\rm d} \Omega^{(d-3)} {\rm d} \Omega^{(d-3)}}{2^{5+2\ep} (2\pi)^{2d-2}}
{\rm d} \varphi_3 \left [ \sin^2(\varphi_3) \right ]^{-\ep} 
\\
&  
\times 
\left [ \xi_1 \xi_2  \right ]^{1-2\ep}
\left [\eta_3 ( 1 - \eta_3 ) \right ]^{-\ep} 
\left [\eta_4 ( 1 - \eta_4 ) \right ]^{-\ep} 
\left [ \lambda (1 - \lambda ) \right ]^{-1/2-\ep}
\frac{|\eta_3 - \eta_4|^{1-2\ep}}{D^{1-2\ep}}
\\
& \times
\left ( 2 E_{\rm max} \right )^{4-4\ep} \theta(\xi_1 - \xi_2)
\theta\left (x_{\rm max}  - \xi_2  \right ) 
\; {\rm d} \xi_1 {\rm d} \xi_2 {\rm d} \eta_3 {\rm d} \eta_4
{\rm d} \lambda.
\end{split} 
\label{eq_337}
\ee
Various variables 
introduced in the above formula parametrize energies and angles of 
the (potentially) unresolved gluons. We use 
\be
E_{3,4} = E_{\rm max} \xi_{1,2},
\;\;\;\;\;\;\;
x_{\rm max} = {\rm min} 
\left [ 1, \frac{1-\xi_1}{\xi_1(1- (1-p_W^2/m_t^2) \xi_1 \eta_{34})} \right ],
\ee
and 
\be
\begin{split} 
& \eta_{34} = \frac{|\eta_3 - \eta_4|^2}{D(\eta_3,\eta_4,\lambda)},\;\;\;\;\;
\sin^2 \varphi_{4} = 4 \lambda (1-\lambda) \frac{|\eta_3 - \eta_4|^2}{
D(\eta_3,\eta_4,\lambda)^2},
\\
& D(\eta_3,\eta_4,\lambda) = \eta_3 + \eta_4 - 2\eta_3 \eta_4 
+ 2 ( 2\lambda - 1) \sqrt{\eta_3 \eta_4 (1-\eta_3) (1-\eta_4)}.
\end{split} 
\ee
The $\eta$-variables are defined as the following scalar products
\be
2 \eta_{3,4} = 1 - \vec n_{3,4} \cdot \vec n_{b},\;\;\;\;\;\;
2 \eta_{34} = 1 - \vec n_{3} \cdot \vec n_{4}.
\ee
Parametrization of the singular phase-space in  Eq.(\ref{eq_337}) 
is still too complicated 
to extract all  singularities; further decomposition is required.  This is 
achieved by a sequence 
of variable changes  that we describe below 
following Refs.~\cite{Czakon:2010td,Czakon:2011ve}.  Specifically, 
we split the phase-space into a sum of five different sectors. 
In each of these sectors, the relation between ``old'' variables 
$\{\eta,\xi \}$ and  ``new'' variables $\{x\}$ reads
\be
\begin{split} 
& {\rm sector~1}:\;\; \xi_1 = x_1,\; \xi_2 = x_1 x_{\rm max} x_2,\; 
\eta_3 = x_3,\;\; \eta_4 = \frac{x_3 x_4}{2},
\\
& {\rm sector~2}:\;\; \xi_1 = x_1,\; \xi_2 = x_1 x_{\rm max} x_2,\; 
\eta_3 = x_3,\;\; \eta_4 = x_3 \left ( 1- \frac{x_4}{2} \right ),
\\
& {\rm sector~3}:\;\; \xi_1 = x_1,\; \xi_2 = x_1 x_{\rm max} x_2 x_4,\; 
\eta_3 = \frac{x_3 x_4}{2} ,\;\; \eta_4 = x_3, \\
& {\rm sector~4}:\;\; \xi_1 = x_1,\; \xi_2 = x_1 x_{\rm max} x_2,\;
\eta_3 = \frac{x_3 x_4 x_2}{2},\;\; \eta_4 = x_3, \\
& {\rm sector~5}:\;\;  \xi_1 = x_1,\; \xi_2 = x_1 x_{\rm max} x_2,\; 
\eta_3 = x_3 \left (1 - \frac{x_4}{2} \right ) ,\;\; 
\eta_4 = x_3.
\end{split} 
\ee
We also write $\lambda = \sin^2(\pi x_5 /2)$. 
This change of variables introduces  
a factor of $\pi$ in the
normalization of the phase-space that is included in the 
expressions below.  The integration region for $x_5$  is 
$x_5 \in [0,1]$.

The full phase-space for $t \to b + g + g + e^+ + \nu$
is given by the sum of five (sector) phase-spaces 
\be
{\rm d} \Pi_{bg_3 g_4 e^+ \nu}  
= \sum \limits_{i=1}^{5} {\rm d} \Pi^{(i)}_{bg_3 g_4 e^+ \nu}.
\ee
Each of the five phase-spaces is written as
\be
{\rm d} \Pi^{(i)}_{bg_3 g_4 e^+ \nu}  
= {\rm Norm} \times {\rm  PS}_{w,i} {\rm PS}_{i}^{-\ep} 
\prod \limits_{k=5}^{9} {\rm d}x_k
\times \prod \limits_{j=1}^{4}  \frac{{\rm d} x_j}{x_j^{1+a_{j}^{(i)} \ep}}
\times \left [ x_1^{b_1^{(i)}} x_2^{b_2^{(i)}} x_3^{b_3^{(i)}} x_4^{b_4^{(i)}} \right ].
\ee
Below we present the functions 
${\rm PS}_{w,i}$, ${\rm PS}_i$ and the exponents $a_{j=1...4}^{(i)}$ and 
$b_{j=1...4}^{(i)}$ for each of the sectors. We note that the normalization factor is common to  all sectors; it reads 
\be
{\rm Norm} = \Upsilon(\ep) 
\left [ \frac{\Gamma(1+\ep)}{(4\pi)^{-\ep} 8 \pi^2}  \right ]^2 
N_{\rm BW} \lp\frac{\Gamma(1-\ep)}{\Gamma(1+\ep)\Gamma(1-2\ep)}\rp^2.
\ee
We also note that we can write 
\be
\begin{split} 
{\rm PS}_{w,i} &= \frac{1}{8\pi^2} 
\frac{E_b E_{\rm max}^4 x_{\rm max}^2}{Q_0 - \vec Q \cdot \vec n_b}\; {\overline {\rm PS}}_{w,i},
%\\
\;\;\;\;\;\;{\rm PS}_i = 
{1024 E_b^2  E_{\rm max}^2 \lambda ( 1-\lambda)
x_{\rm max}^2
(1-x_3)  }
\sin^2\left ( \varphi_{3} \right ) \; {\overline {\rm PS}_i}.
\label{eq352}
\end{split} 
\ee
The expressions for the exponents and the phase-space factors for each of the 
five  sectors read\footnote{We suppress the sector label 
everywhere in the equations below.}
\be
\begin{split}
 {\rm \underline{sector}~1}&: \;\; 
\{ a_1 = 4, a_2 = 2, a_3 = 2, a_4 = 1 \}, \;\;\;\;
\{b_1 = 4, b_2 = 2, b_3 = 2, b_4 = 1\}, \\
& {\overline {\rm PS}}_w = \frac{(1-\frac{x_4}{2})}{
2N_1(x_3,\frac{x_4}{2},\lambda)},\;\;\;{\overline {\rm PS}} = \frac{
\left (1-\frac{x_3x_4}{2}  \right )
\left ( 1- \frac{x_4}{2} \right )^2
}{2 N_1^2(x_3,\frac{x_4}{2},\lambda)},
\\
  {\rm \underline{sector}~2}&:\;\;
 \{ a_1 = 4, a_2 = 2, a_3 = 2, a_4 = 2 \}, \;\;\;\;
\{b_1 = 4, b_2 = 2, b_3 = 2, b_4 = 2\}, \\
& {\overline {\rm PS}}_w = \frac{1}{4N_1(x_3,1-\frac{x_4}{2},\lambda)},
\;\;\;
{\overline {\rm PS}} = 
\frac{\left (1-\frac{x_4}{2}  \right )
\left ( 1-x_3(1-\frac{x_4}{2}) \right )
}{4 N_1^2(x_3,1-\frac{x_4}{2},\lambda)}, 
\\
 {\rm \underline{sector}~3} & :\;\;
 \{ a_1 = 4, a_2 = 2, a_3 = 2, a_4 = 3 \}, \;\;\;\;
\{b_1 = 4, b_2 = 2, b_3 = 2, b_4 = 3\}, \\
& {\overline {\rm PS}}_w = 
\frac{(1-\frac{x_4}{2})}{2
N_1(x_3,\frac{x_4}{2},\lambda)},
\;\;\;\;
{\overline {\rm PS}} = \frac{ \left (1 - \frac{x_3 x_4}{2} \right ) 
\left (1-\frac{x_4}{2}  \right )^2
}{2 N_1^2(x_3,\frac{x_4}{2},\lambda)},
\\
 {\rm \underline{sector}~4}& :\;\;
 \{ a_1 = 4, a_2 = 3, a_3 = 2, a_4 = 1 \}, \;\;\;\;
\{b_1 = 4, b_2 = 3, b_3 = 2, b_4 = 1\}, \\
& {\overline {\rm PS}}_w = 
\frac{(1-\frac{x_2x_4}{2} )}{2
 N_1(x_3,\frac{x_4 x_2}{2},\lambda)},
\;\;\;\;
{\overline {\rm PS}} = \frac{
(1-\frac{x_2x_3x_4}{2}) \left (1 - \frac{x_2 x_4}{2} \right )^2 
}{2 N_1^2(x_3,\frac{x_4 x_2}{2},\lambda)} 
,
\\
 {\rm \underline{sector}~5}&:\;\;
 \{ a_1 = 4, a_2 = 2, a_3 = 2, a_4 = 2 \}, \;\;\;\;
\{b_1 = 4, b_2 = 2, b_3 = 2, b_4 = 2\}, \\
& {\overline {\rm PS}}_w = 
\frac{1}{4
N_1(x_3,1-\frac{x_4}{2},\lambda)},
\;\;\;
{\overline {\rm PS}} = \frac{
( 1-\frac{x_4}{2} ) ( 1-x_3(1-\frac{x_4}{2} ) )
}{4 N_1^2(x_3,1-\frac{x_4}{2},\lambda)}.
\end{split}
\ee
The function $N_1$ is given by 
\be
        N_1(x_3,x_4,\lambda) = 1 + x_4 (1 -2 x_3)  
        - 2(1-2 \lambda)\sqrt{x_4(1-x_3)(1-x_3 x_4)}.
\ee
It is straightforward to use momentum and  
phase-space parametrizations described 
above  to extract singularities from matrix elements squared 
in a factorized 
form, thereby facilitating the integration over the full phase-space with 
the help of plus-distribution expansions.  We will briefly discuss 
this in the next Section. 

We note that  the phase-space parametrization that we just described 
resolves all the singularities of  the matrix elements for the 
final state 
with a $b$-quark and two gluons. The process with  
a $b$-quark and two additional massless quarks in the final state 
has a simpler singularity structure than the two-gluon 
case, so our phase-space parametrization should be 
applicable to the quark case as well.  However, the quark case is 
different in that quark amplitudes are, in general, not symmetric 
with respect to $q$ and $\bar q$ permutation 
and also they are less singular. It is therefore useful 
to consider $b q \bar q$ final state 
 separately and design a  parametrization of 
the phase-space for this case. We start by listing kinematic configurations 
that lead to non-integrable singularities in the amplitude 
$ t \to b + q(p_3) + \bar q(p_4) + e^+ + \nu$.  
From the point of view of the singularity structure it is sufficient 
to understand  the 
singlet contribution  where the flavor of the 
$q \bar q$-pair is different from 
$b$.  The case  of $b b \bar b$ final state then easily follows. 
The singular  configurations that we need to consider are 
{\it i}) both $q$ and $\bar q$ are soft; 
{\it ii}) both $q$ and $\bar q$ are collinear to the $b-$quark 
and {\it iii}) $q$ and $\bar q$ are collinear to each other. 
Analyzing these singular limits, we  conclude that 
the number of sectors that we need to consider for 
$b q \bar q$ final state is four. Momenta are parametrized as in 
the gluon case. The first two sectors are 
\begin{itemize}
\item sector 1: $E_3 = x_1 E_{\rm max}$, \quad $E_4 = x_1 x_2 x_{\rm max} E_{\rm max}$, \quad $\eta_3 = x_3$, \quad $\eta_4 = x_3(1-x_4)$,
\item sector 2: $E_3 = x_1 E_{\rm max}$, \quad $E_4 = x_1 x_2 x_{\rm max} E_{\rm max}$, \quad $\eta_3 = x_3(1-x_4)$, \quad $\eta_4 = x_3$,
\end{itemize} 
and the other two are obtained 
from sectors 1 and 2  by permuting $p_3$ and $p_4$.
Finally, all quark sectors share the same phase-space. It reads 
\be
\begin{split} 
& {\rm d}\Pi_{b q\bar qe^+ \nu} = 
 \Upsilon(\ep) 
\left[\frac{\Gamma(1+\ep)}{(4\pi)^{-\ep}} \frac{1}{8\pi^2}\right]^2
N_{\rm BW} \lp\frac{\Gamma(1-\ep)}{\Gamma(1+\ep)\Gamma(1-2\ep)}\rp^2
\\
&\times \frac{1}{8\pi^2}\frac{E_b E_{\rm max}^4 x_{\rm max}^2  }{(Q_0-Q_3)}
\frac{x_2}{N_1(x_3,1-x_4,\lambda)}
{\rm PS}^{-\ep}
\\
&  
\times 
\frac{{\rm d} x_1}{x_1^{1+4\ep}} \frac{{\rm d} x_3}{x_3^{1+2\ep}}
\frac{{\rm d} x_4}{x_4^{1+2\ep}} {\rm d} x_2 {\rm d} x_5 {\rm d} 
x_6 {\rm d} x_7 {\rm d} x_8 {\rm d} x_9
[x_1^4 x_3^2 x_4^2] ,
\end{split}
\ee
where 
\be
{\rm PS}
=  \frac{ 1024 E_b^2 E_{\rm max}^2
x_2^2 x_{\rm max}^2 
}{N_1^2(x_3,1-x_4,\lambda)
} 
\lambda(1-\lambda) (1-x_3)(1-x_4)(1-x_3 (1-x_4))\sin^2\varphi_3.
\ee
It follows from the phase-space parametrization that 
we do not intend to construct the plus-distribution expansion  
with respect to  the variable $x_2$. This is done intentionally since 
$x_2 \to 0$ corresponds to a single-soft limit which is not singular for 
the $q \bar q b$ final state.

\section{Singular limits}
\label{limits}

The phase-space parametrization that we discussed in the 
previous Section is optimized 
for extracting the singular behavior of  scattering amplitudes. This 
is done by exploiting universal soft and collinear limits of these 
amplitudes. The goal of this Section is to describe the limits that 
we require for the  top quark decay. We begin with the discussion of the 
relevant soft limits and then consider the collinear ones. 

Soft singularities can only occur 
if gluons become soft.\footnote{In the process
$ t \to b + q + \bar q + e^+ + \nu$ there is also 
a soft singularity when energies of {\it both} $q$ and $\bar q$ 
vanish. We do not  discuss this case as it follows from the soft 
limit of $ t \to b + g + e^+ + \nu$.} 
In general, emissions of soft particles 
lead to color correlations but because in the top quark decay 
the number of colored particles is small, all color correlations 
simplify and  the relevant soft limits 
can be written in a compact form.  For example, the 
single soft limit of the $t \to b + g(p_3) + e^+ + \nu$ amplitude squared  
reads  
\be
|{\cal M}(t,b,g)|^2 \approx g_s^2 C_F |{\cal M}(t,b)|^2 S_{tb}(p_3),
\label{esl}
\ee
where the eikonal factor $S_{tb}$ is 
\be
S_{tb}(p_3) = \frac{2 p_t \cdot p_b}{( p_t \cdot p_3) ( p_b \cdot p_3) }
- \frac{m_t^2}{(p_t \cdot p_3)^2 }.
\ee
Note that for 
a single-gluon emission process $t \to b + g(p_3) + e^+ + \nu$ the 
function $F(x_1,x_2)$ is given by 
\be
F(x_1,x_2) = x_1^2 x_2  |{\cal M}(t,b,g)|^2,
\ee
where $x_1$ and $x_2$ parametrize the gluon energy $E_3 = E_{\rm max} x_1$ 
and the relative angle with respect to the $b$-momentum, 
$ p_b \cdot p_3 =2 E_b E_3 x_2$. We have seen that we need $F(x_1=0,x_2)$ 
for the computation and we now show how to obtain it from the 
soft limit Eq.(\ref{esl}). The key point is that the dependence on the 
gluon momentum in the right hand side of Eq.(\ref{esl}) is entirely contained 
in the eikonal factor. Hence, to find $F(x_1=0,x_2)$, we need 
\be
\lim_{x_1 \to 0}^{} x_1^2 x_2 S_{tb}(p_3) 
= E_{\rm max}^{-2} ( 1- x_2).
\ee
While the single soft limit provides a very simple example, we note that
the same approach is  used to compute  
limiting values of integrand functions $F$ 
also for more complex cases, such as limits of double-real emission 
amplitudes squared and limits of 
one-loop contributions  to single-real emission processes.  We will not 
discuss these derivations; instead,
we limit ourselves to the  presentation of the relevant ingredients 
in what follows.

The double soft limit of the amplitude squared for 
$t \to b + g(p_3) + g(p_4) + e^+ + \nu$ is only slightly more complicated
than the single soft limit.  We can write it as 
\be\label{doublesoft}
\begin{split} 
 |{\cal M}(t,b,g_3,g_4)|^2 \approx 
& g_s^4 |{\cal M}(t,b)|^2 C_F
\left (C_A ( 2 S_{tb}(p_3,p_4) - S_{tt}(p_3,p_4) - S_{bb}(p_3,p_4) )
\right.\\
& \left. 
 + C_F S_{tb}(p_3) S_{tb}(p_4) \right ). 
\end{split} 
\ee
The double-emission eikonal factors 
depend on the masses of the emitters; they can be extracted 
from Refs.\cite{Catani:1999ss,Czakon:2011ve}.
For completeness, 
we present them 
below for the specific cases that we need  to describe 
 top quark decays. 
In the massless
case $p_b^2 = 0$, we find 
\be
S_{bb}(p_3,p_4) =   2(1-\ep)\frac{
\lp p_b \cdot p_3\rp \lp p_b \cdot p_4\rp}
{(p_3 \cdot p_4)^2 (p_b \cdot p_{34})^2}, 
\ee
where $p_{34} = p_3 + p_4$.  In the massive case $p_t^2 = m_t^2$, the double eikonal factor reads 
\be
\begin{split}
& S_{tt}(p_3,p_4) = 2(1-\ep)\frac{\lp p_t \cdot p_3\rp \lp  p_t \cdot p_4 \rp}
{ \lp p_3\cdot p_4\rp^2 \left ( p_t \cdot p_{34} \right )^2 }
+ \frac{m_t^4}{\lp p_t\cdot p_3\rp \lp p_t\cdot p_4\rp \lp p_t\cdot p_{34}\rp^2} +
\\
& + \frac{3}{2} \frac{m_t^2}{\lp p_3\cdot p_4\rp \lp p_t\cdot p_3\rp
\lp p_t\cdot p_4\rp}
-\frac{m_t^2}{\lp p_3\cdot p_4\rp \lp p_t \cdot p_{34}\rp^2}
\lp \frac{p_t\cdot p_3}{p_t\cdot p_4} + \frac{p_t\cdot p_4}{p_t\cdot p_3} +4\rp.
\end{split}
\ee
Finally, we give an expression for the eikonal factor $S_{tb}(p_3,p_4)$, where $p_t^2 = m_t^2$ 
and $p_b^2 = 0$. It reads 
\be
\begin{split}
& S_{tb}(p_3,p_4) =  
(1-\ep) \frac{\lp p_b\cdot p_4\rp \lp p_t \cdot p_3\rp + 
\lp p_b\cdot p_3\rp \lp p_t \cdot p_4\rp}{\lp p_3\cdot p_4\rp^2
\lp p_b\cdot p_{34}\rp \lp p_t\cdot p_{34}\rp} + 
\frac {p_b\cdot p_t}{p_3\cdot p_4} \lp
\frac{1}{\lp p_b \cdot p_3\rp \lp p_t\cdot p_4\rp} + \right.\\
& \left. +\frac{1}{\lp p_b \cdot p_4\rp \lp p_t\cdot p_3\rp}
-\frac{1}{\lp p_b \cdot p_{34} \rp \lp p_t \cdot p_{34} \rp }
\lp 3 + \frac{\lp p_b\cdot p_4\rp \lp p_t\cdot p_3\rp }
{2 \lp p_b\cdot p_3\rp \lp p_t\cdot p_4\rp } +
\frac{\lp p_b\cdot p_3\rp \lp p_t\cdot p_4\rp }
{2 \lp p_b\cdot p_4\rp \lp p_t\cdot p_3\rp }\rp\rp + \\
& - \frac{\lp p_b\cdot p_t\rp^2}{\lp p_b\cdot p_3\rp \lp p_b\cdot p_4\rp
\lp p_t\cdot p_3\rp \lp p_t\cdot p_4\rp}
\lp 1 - \frac{\lp p_b\cdot p_3\rp \lp p_t\cdot p_4\rp + 
\lp p_b\cdot p_4\rp \lp p_t\cdot p_3\rp}{2 \lp p_b\cdot p_{34}\rp \lp 
p_t\cdot p_{34}\rp}\rp + \\
& + m_t^2 \lp
\frac{\lp p_b \cdot p_t\rp}
{2 \lp p_b\cdot p_3\rp \lp p_b\cdot p_4\rp
\lp p_t\cdot p_3\rp \lp p_t\cdot p_4\rp}\frac{p_b\cdot p_{34}}{p_t\cdot p_{34}}
-\frac{1}{4 \lp p_3\cdot p_4\rp \lp p_t \cdot p_3\rp \lp p_t \cdot p_4\rp} + 
\right.\\
&\left. -\frac{1}{2 \lp p_3\cdot p_4\rp \lp p_b\cdot p_{34}\rp 
\lp p_t \cdot p_{34}\rp} \lp
\frac{\lp p_b \cdot p_3\rp^2}{\lp p_b\cdot p_4\rp \lp p_t\cdot p_3\rp}
+\frac{\lp p_b \cdot p_4\rp^2}{\lp p_b\cdot p_3\rp \lp p_t\cdot p_4\rp}
\rp
\rp.
\end{split}
\ee       
It is straightforward to check,  using momenta parametrization  
in terms of $x$-variables for various types of 
sectors for double-real emission described 
in Section~\ref{phasespace}, 
that all overlapping singularities get resolved in the 
sum of the eikonal factors
Eq.(\ref{doublesoft}) and that the subtraction terms for 
the soft limits of the double-real emission process can be calculated.

%\subsection{The Soft Current at One-Loop Order}

In order to subtract soft singularities 
from one-loop amplitudes for $t \to b +g(p_3) + e^+ + \nu$, we need 
the expression for the eikonal  
current to  one-loop. The massless result  for the one-loop soft current 
was obtained in Ref.~\cite{catanisoft}. The one-loop massive soft current was computed in 
Ref.~\cite{Bierenbaum:2011gg}. We have re-calculated 
this soft current for 
the top quark 
decay into a bottom quark and present the result of the calculation below. 
This result is identical to Ref.~\cite{Bierenbaum:2011gg} but 
it is written in the kinematic 
region which is relevant for the top quark decay, so that 
no analytic continuation is required.  The soft limit of the 
interference of one-loop and tree-level  amplitudes is 
written in Eq.(\ref{eq_rv2}). The one-loop eikonal factor reads 
\be
S^{(1)}_{tb}(p_3) = g_s^2 C_F S_{tb}(p_3) 2 {\rm Re} 
\left [ {\cal I}_{tb}(p_t,p_b,p_3) \right ],
\ee
where 
\be
\begin{split} 
   {\cal I}_{tb}  
= & - \frac{(p_t \cdot p_b)}{m_t^2 (p_b \cdot p_3)^2-2 (p_t \cdot p_b) (p_t \cdot p_3) (p_b \cdot p_3)} \Bigg \lbrace  2 (p_t \cdot p_3)^2 (p_b \cdot p_3) 
 I_1
\\  
&
  +   2 (p_t \cdot p_3) (p_t \cdot p_b) (p_b \cdot p_3) I_2
+   2 (p_b \cdot p_3) \left[ (p_t \cdot p_b) (p_t \cdot p_3)-m_t^2 (p_b \cdot p_3)\right] I_3
\\
 & +   
4  \left[ (p_t \cdot p_b) (p_t \cdot p_3) (p_b \cdot p_3)
-m_t^2 (p_b \cdot p_3)^2 \right] \frac{(p_t 
 \cdot p_3) (p_b \cdot p_3)}{(p_t \cdot p_b)} I_4 \Bigg\rbrace .
\end{split} 
\label{softI}
\ee
The Feynman integrals  $I_{1,..4}$ in Eq.(\ref{softI}) 
are defined as 
\bea
\label{masterintegrals}
&  & I_1 = \int \frac{d^d k}{i (2 \pi)^d} \frac{1}{(k+p_3)^2 k^2 (2 p_t 
\cdot k)}, 
\;\;\;\;\;\;\;\;
  I_2 =  \int \frac{d^d k}{i (2 \pi)^d} \frac{1}{k^2 (2 p_t \cdot k) 
(2 p_b \cdot k + 2 p_b \cdot p_3)},
\\
&&  I_3 =  \int \frac{d^d k}{i (2 \pi)^d} \frac{1}{k^2 (2 p_b \cdot k) (
2 p_t \cdot k - 2 p_t \cdot p_3)},
\;\;\;\;
I_4 = \int \frac{d^d k}{i (2 \pi)^d} \frac{1}{(k-p_3)^2 k^2 (2 p_b 
\cdot k) (2 p_t \cdot k - 2 p_t \cdot p_3)}.
\nonumber 
\eea
We note that each propagator 
implicitly includes $+i0$-prescription that unambiguously specifies 
 the relevant 
analytic continuation of the result.   The integrals are computed to be (see Appendix~\ref{oneloopMI} for details)
\bea
&&  I_1 =  \frac{\Gamma(-\epsilon) \Gamma(2 \epsilon) m_t^{2 \epsilon}}{ (4 \pi)^{d/2} 
(s_{t3})^{1+2 \epsilon}
},\;\;\;\;\;\;
I_2 =  -\frac{ \Gamma(1-2 \epsilon) \Gamma(\epsilon) \Gamma(2 \epsilon) 
m_t^{-2 \epsilon} e^{2 \pi i \epsilon}
}{(4 \pi)^{d/2}
(s_{b3})^{2 \epsilon}
(s_{tb})^{1-2 \epsilon}},\;\;\;
I_3 =  -\frac{\Gamma(-\epsilon) \Gamma(2 \epsilon) m_t^{2 \epsilon} }{
(4\pi)^{d/2}
(s_{t3})^{2 \epsilon} 
s_{tb} }, \nonumber \\
&& I_4 =  -\frac{ \Gamma(-\epsilon) \Gamma(1+2 \epsilon) m_t^{2 \epsilon} }{(4 \pi)^{d/2}  
(x_2)^{\epsilon} (s_{t3})^{1+2 \epsilon} s_{b3}} \left[ e^{2 \pi i  \epsilon} 
\frac{\Gamma(1+\epsilon) \Gamma(-2 \epsilon)}{\Gamma(1-\epsilon)} 
F_{21} (-\epsilon, 1 + \epsilon, 1- \epsilon ; 1 - x_2) 
\right. 
\label{eq_me}
\\
& & \left. 
~~~~~~~~~~~~- \frac{1}{\epsilon}  F_{21} (-\epsilon, -\epsilon, 1- \epsilon ; 1 - x_2) \right],
\nonumber 
\eea
where $s_{ij} = 2 p_i \cdot p_j$ 
and 
$x_2 = ( m_t^2 s_{b3})/( s_{tb} s_{t3} )$. We emphasize  that 
products of all 
four-momenta are positive-definite 
 since we write the results for the soft current 
in the correct kinematic region. We note that to use the results Eq.(\ref{eq_me}) in our plus-distribution
expansion, we should extract all the relevant branch-cuts from the hypergeometric functions. 

We now turn to the discussion of collinear singularities. 
It is well-known that in the collinear limit  
any  matrix element squared 
factorizes into a splitting function that is,  in general, 
spin-dependent, and a reduced squared matrix element
that depends only on the sum of momenta of collinear 
particles.  
For the top quark decay, we need to describe such collinear splitting processes 
as 
$ b \to b + g$, $b \to b + g + g$, $g \to g + g$,
$b \to b + q + \bar q$, $b \to b + b \bar b$ and $ g \to q \bar q$.
  The corresponding 
splitting functions can be found in 
Ref.~\cite{Catani:1999ss}.   Many  splitting functions 
that we need for top quark decay  do not contain spin-correlations.  The 
corresponding factorization formula reads 
\be
|{\cal M}|^2 \sim s^{-1}_c \; P(\{z \})
 |{\tilde {\cal M}}|^2,
\ee
where ${\tilde {\cal M}}$ is the reduced amplitude,  
$s_c$ is the scalar product of collinear momenta that 
vanishes in the exact collinear limit and $P(\{z \})$ is a splitting 
function that depends on the set of energy fractions $\{ z \}$.
This is an easy case to deal with since, in this situation, the 
collinear limit depends on the reduced amplitude 
squared which can be computed in the standard way.  
However, there are two splitting functions that we have to deal with 
($g \to gg$, $g \to q \bar q$),  that 
do contain spin correlations. 
In that case, the collinear limit is more complicated
\be
|{\cal M}|^2 \sim s^{-1}_c \;
P^{\mu \nu}(\{z \} )
{\tilde {\cal M}}^\mu {\tilde {\cal M}}^{\nu,*}.
\label{eq456}
\ee
Any  spin-dependent splitting function $P^{\mu \nu}(z)$ can
be written as 
\be
P(z) = -g^{\mu \nu} P_1(z,\ep) + n^\mu n^\nu P_2(z,\ep),
\label{eq457}
\ee
where $n^\mu$ is a four-vector that specifies how the collinear 
direction is approached.  To compute the right hand side in Eq.(\ref{eq456}), 
it is useful to re-write it through helicity amplitudes.  
This is  easy  to do since  
gauge-invariance of scattering amplitudes allows us to replace 
$g^{\mu \nu}$ and $n^\mu n^\nu$ in  Eq.(\ref{eq457}) by
rank-two tensors constructed from   polarization vectors
\be
-g_{\mu \nu} = \sum \limits_{\lambda= \pm} 
\epsilon_{\mu}(\lambda) \epsilon_{\nu}^{ *}(\lambda),\;\;\;
n^{\mu} n^{\nu} = 
\sum \limits_{\lambda_1=\pm,\lambda_2 = \pm} 
[ n \cdot \epsilon(\lambda_1) \;
n \cdot \epsilon(\lambda_2) ]\;  
\epsilon_{\mu}^*(\lambda_1) 
\epsilon_{\nu}^*(\lambda_2). 
\label{pol_sum}
\ee
We note that, since we work in dimensional regularization,
it is important  to extend the  sum 
over four-dimensional gluon helicities 
in Eq.(\ref{pol_sum}), 
to include  $\ep$-dimensional polarization vectors as well. 
It is possible to do so by 
introducing ``scalar'' polarizations of gluons that point along 
Cartesian components of the $\ep$-dimensional sub-space  of the full $d$-dimensional 
Minkowski space. 
 In Section~\ref{correct} 
we will return  to the issue 
of spin-correlations, to discuss a consistent implementation of angular 
variables in our phase-space parametrization.

While the majority of collinear 
limits is completely determined by the known 
splitting functions~\cite{Catani:1999ss},
there are a few cases when further computations 
are required.  This happens for example in the strongly-ordered  
limit of the $b \rightarrow g(p_3) g(p_4) b$ splitting function. 
The strongly-ordered limit occurs when the two gluons are 
much more collinear  to each other than to the 
$b$-quark, i.e.  $ p_3 \cdot p_4   \ll p_b \cdot p_3 + p_b \cdot p_4$.
In this case the full matrix element factorizes as
\be
 |{\cal M}(t,b,g_3,g_4)|^2 \approx \frac{g_s^2}{(2 p_3 \cdot p_4)  
( 2 p_b \cdot p_3 + 2 p_b \cdot p_4 ) } 
\hat{P}^{\rm s.o.}_{g_3 g_4 b} |{\cal M}(t,b)|^2 .
\ee
We derive the following result for the 
strongly-ordered limit of the $b \to g(p_3) g(p_4) b$ splitting 
function 
\be 
\begin{split}
\hat{P}^{\rm s.o.}_{g_3 g_4 b} 
= & 8 C_F C_A  \left[ \frac{4 \lambda  z_3 (1-z_3)  z_b}{1-z_b} 
+ \frac{1 + (z_3 (1-z_3)(1-z_b))^2  + z_b^2 
- 2  z_3 (1-z_3)  (1+z_b^2)}{(1-z_b)z_3 (1-z_3)} \right] 
\\
&+8 C_F C_A \left[(1-z_b)  \left(2 
- z_3 (1-z_3) - \frac{1}{z_3 (1-z_3)} \right)  
- \frac{ 4 \lambda z_3 (1-z_3) z_b }{ 1-z_b}  
\right] \epsilon + \mathcal{O}(\epsilon^3),
\end{split} 
\ee
where $z_3 = E_{3}/(E_{3}+E_{4})$,  
$z_b = E_b/(E_{3} + E_{4}+E_b)$ and $\lambda$ is one of the 
phase-space variables introduced 
in Section~\ref{phasespace}.

\section{How to treat spin-correlations consistently in the NNLO computation}
\label{correct}

As we pointed out earlier, it is plausible, but not entirely obvious, 
that one can consistently  
treat  positron and neutrino momenta as four-dimensional 
in higher-order perturbative computations. The goal of this 
Section is to elaborate on this point.  We will discuss it 
in a slightly simplified  setting, by considering top quark decay 
into a polarized $W$-boson, ignoring $W$-decays into leptons. 

We want to treat the polarization vector of the 
$W$ -boson as a four-dimensional vector. In addition, there are 
other vectors that we would like to treat 
as four-dimensional; they include the $W$-momentum, 
the top quark momentum and the $b$-quark momentum.  
Assuming that we are in the 
top quark rest frame and making use of momentum conservation, we conclude 
that we can embed the momentum of one additional gluon to the four-dimensional 
space, but that the second gluon must have additional $\ep$-dimensional 
components.  This means that the  phase-space parametrization 
for single-gluon  emission processes 
that we discussed in 
Section~\ref{phasespace} is perfectly valid but  the 
phase-space parametrization for 
double-real emission computations must be extended. 
To this end, we parametrize momenta and the $W$-polarization vector 
as 
\be
\begin{split}
& p_t = (m_t, \vec 0;0),\; p_b = E_b(1,\vec 0,1;0),
\;p_W = (E_W,p_w^x,0,p_W^y;0),
\;\epsilon_W = (\epsilon_W^{(0)},\epsilon_W^{(x)},\epsilon_W^{(y)},
\epsilon_W^{(z)};0),
\\
& 
p_{g_3} = E_{g_3} 
\left (1, \sin \theta_3 \cos \varphi_3, \sin \theta_3 \sin \varphi_3 \cos \alpha,
\cos \theta_3 ; \sin \theta_3 \sin \varphi_3 \sin \alpha \right ),
\label{split}
\end{split}
\ee
and determine $p_{g_4}$ using  momentum conservation.  A component 
of a four-vector in Eq.(\ref{split}) 
shown after the semi-column is the $\ep$-dimensional 
component.  We now perform 
a rotation in the $y-\ep$ plane, to remove 
the $\ep$-dimensional part of the gluon momentum $p_{g_3}$.  From 
momentum parametrization of other particles it follows that this rotation 
does not induce  $\ep$-dimensional components for any of them. 
On the other hand, such $\ep$-components appear 
in the polarization vector $\epsilon_W$ of the $W$-boson after the 
rotation.

To understand in which cases we require the $\ep$-dimensional components 
in $W$-polarization,  we consider  
singular limits. These limits can be divided into two 
groups. Some limits lead to expressions where squares of 
reduced matrix elements appear. 
The reduced matrix elements squared can be either leading order 
or next-to-leading order. In both cases, because of the choice 
of the momenta above, we conclude that 
\be
|{\cal M}|_{\rm LO,NLO}^2(p_t,p_b,p_W,\epsilon_W(\alpha)) = 
|{\cal M}|^2_{\rm LO,NLO}(p_t,p_b,p_W,\epsilon_W(0)),
\label{sqrs}
\ee
where $\alpha$ parametrizes the $\ep$-dimensional angle and $\alpha=0$ 
corresponds to the four-dimensional limit, see  Eq.(\ref{split}).  We note 
that Eq.(\ref{sqrs}) is valid because 
through next-to-leading order 
 none of the vectors contains  $y$-components that can 
lead to an $\alpha$-dependence through products with $\epsilon_W(\alpha)$. 
Therefore, in all limits that are proportional to the reduced matrix elements 
squared, we can neglect the dependence of the $W$-polarization vector 
on the extra-dimensional angle $\alpha$, provided that 
the normalization condition $\epsilon_W^2 = -1$ is maintained. 

A different situation occurs when spin correlations appear 
in the singular limits.
In the top quark decay case, 
there are no triple-collinear limits which have 
this feature and we only need to focus on double-collinear limits. As 
an example, we take the case when gluon $g_3$ and gluon $g_4$ are collinear.
The collinear limit is written as 
\be
\label{eq24}
|{\cal M}_{t \to b W g_3g_4}|^2 \sim 
P_{\mu \nu} {\cal M}^{\mu} {\cal M}^\nu
= \left ( -g_{\mu \nu} P_1(z,\ep) +  P_2(z,\ep) n^\mu n^\nu
\right ) {\cal M}^\mu_{t \to b W g_{34}} {\cal M}_{t \to bWg_{34}}^{\nu},
\ee
and $n^\mu$ is the vector that describes how vectors $p_{g_3,g_4}$ approach 
the collinear direction. The matrix element ${\cal M}_{t \to b W g_{34}}^\mu$ 
contains the polarization of the $W$-boson and depends on momenta of $t,b$ 
and $W$. Since none of these momenta have $y$-components, we can 
remove the $\alpha$-dependence from the matrix element by rotating 
all vectors in the $y-\ep$ plane.  Since, in general, the vector $n^\mu$ 
{\it does have} a $y$-component, such a rotation induces an $\alpha$-dependence 
in the splitting function. The vector $n^\mu$ becomes 
\be
\begin{split}
& n^\mu \to 
\pm \sqrt{\lambda} {\bar n}^\mu
%\\
%& 
+ \sqrt{1-\lambda} \left ( \cos \alpha {\tilde n}^\mu
 + \sin \alpha n_5^\mu \right ), 
\end{split}
\ee
where $
{\bar n}^\mu = \left [ \cos \theta_3 \cos \varphi_3, 
\cos \theta_3 \sin \varphi_3, -\sin \theta_3; 0 \right ], 
$
$
\tilde{n}^\mu = \left [ -\sin \varphi_3, \cos \varphi_3 , 0 ; 0 \right ]$ and  $
{n}_5^{\mu} = \left[0,0,0;1 \right ]
$.
The two possible signs in front of the first vector are for different sectors. 
Since we do not need the  $\alpha$-angle  for any other limit, 
it should decouple from the point-by-point subtraction. This 
allows us to average  over  $\alpha$ and modify the spin-correlated 
part of the splitting function by including 
 ``integrated'' terms.  Using the two non-vanishing  averages 
\be
\langle \cos^2 \alpha \rangle = \frac{1}{1- 2\ep},
\;\;\;
\langle \sin^2 \alpha \rangle = \frac{-2\ep}{1-2\ep},
\ee
we find the replacement  rule 
\be
n^\mu n^\nu 
\to \lambda {\bar n}^\mu {\bar n}^\nu
+ \frac{2\ep(1-\lambda)}{1-2\ep} \left ( 
\tilde{n}^\mu \tilde{n}^{\nu} - n_5^{\mu} n_5^{\nu} \right ).
\ee
Applying it to Eq.(\ref{eq24}) and treating all the vectors in the 
matrix element ${\cal M}$, 
including $W$-polarization, as four-dimensional, we get an easy way 
to consistently combine dimensional regularization with four-dimensional 
handling of all the relevant vectors that we have in the process. Finally, 
we note that it is obvious how to generalize this discussion to account for 
the $W$-boson decays to a lepton pair.

\section{Results} 
\label{results}

In this Section we show some results of our computation. 
We note that our goal in this paper is not 
to provide extensive phenomenological studies 
 of top quark decays but to illustrate the capabilities of the 
computational method.  The results that we present below are selected 
accordingly. 

We   remind the reader that we consider the top quark decay  into 
hadronic final states  and a single  
lepton pair $t \to b +X_{\rm hadr} + e^+ + \nu$.   We use the on-shell 
renormalization scheme  for the top quark 
 and the $\overline {\rm MS}$ renormalization scheme 
for the strong coupling constant in a theory with five  massless flavors. 
We set the renormalization scale to the value of the top quark mass 
$\mu = m_t$.  Results for $\mu\ne m_t$ can be immediately obtained by using 
renormalization group equations, so we do not  present them here. 
In our computation, we account exactly  for the  
 off-shellness of  the $W$-boson. The reasons for doing that 
are explained in the Introduction. 

\begin{figure*}
\centering
\includegraphics[scale=0.6]{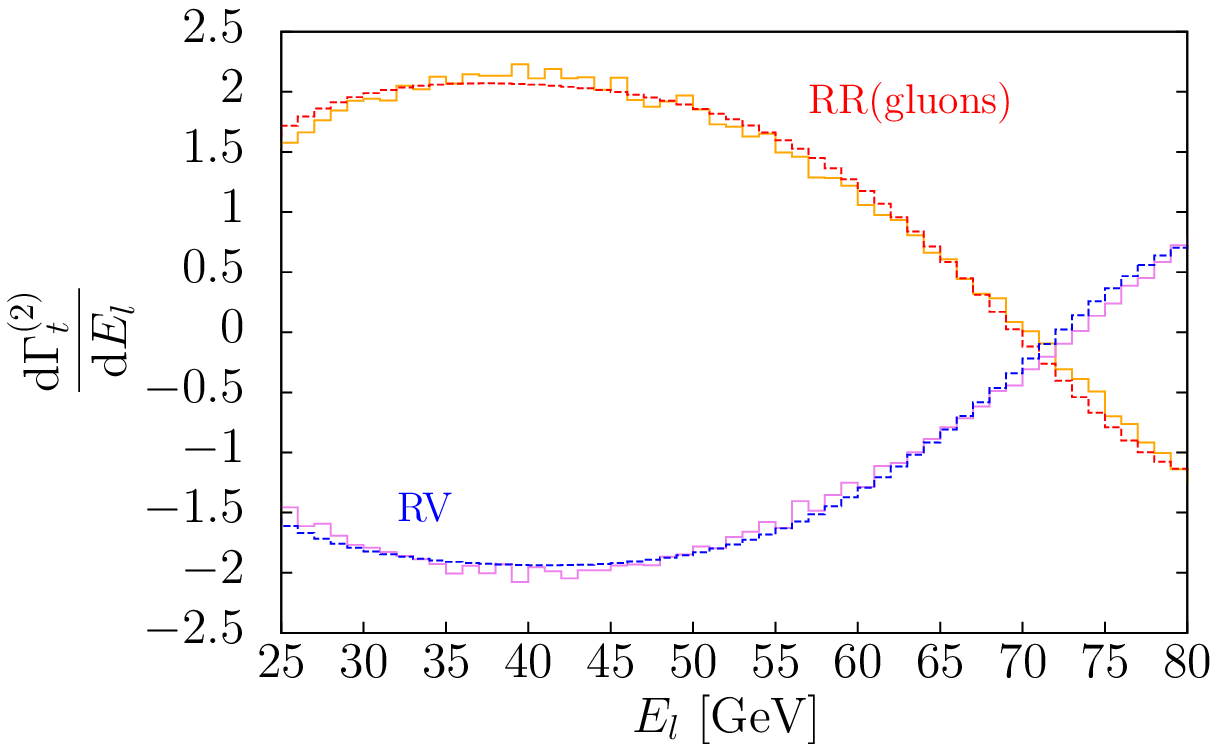}
\includegraphics[scale=0.6]{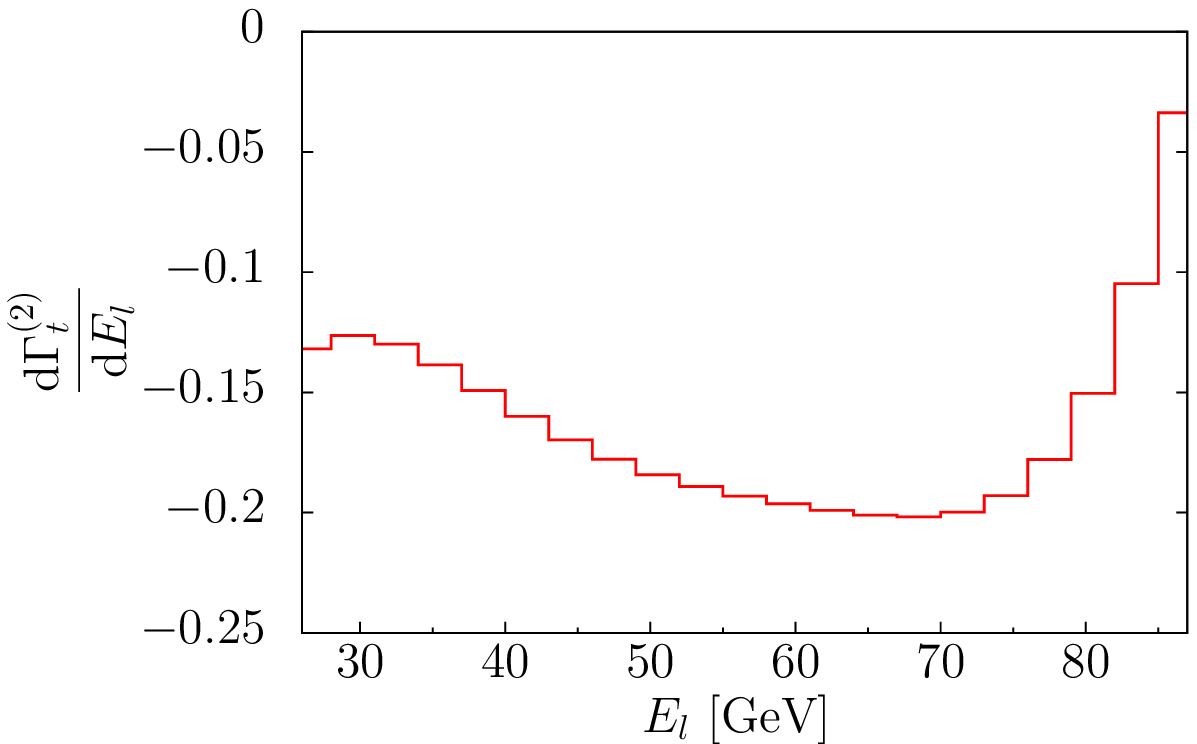}
\caption{ 
Distribution of the second order coefficient ${\rm d} \Gamma_t^{(2)}$ 
in the positron energy. 
Left pane: double-real and 
real-virtual contributions. 
Original histograms are shown with solid lines; Legendre polynomial 
fit is shown with dashed lines. 
Right pane: distribution of 
${\rm d}\Gamma_t^{(2)}$ in the lepton energy.
See text for details.}
\label{enrg}
\end{figure*}

To show the quality of our numerical computation,  we 
isolate all the 
different  color structures that contribute to the final result
\be
\Gamma_t = \Gamma_t^{(0)} \left  [ 
1 + C_F \frac{\alpha_s}{2\pi} \Delta_1 
+ \left ( \frac{\alpha_s}{2\pi} \right )^2
\left ( C_F^2 \Delta_{A} + C_F C_A \Delta_{NA}
+ C_F T_R n_l \Delta_{l} + C_F T_R \Delta_{h} 
\right ) \right ]+ {\cal O}(\alpha_s^3).  
\label{eqr}
\ee
The different contributions to second order corrections were  computed 
in a number of papers a decade ago for top quark decays to an on-shell $W$ boson.  
Since $\Delta_{A, NA, l, h}$, as  defined in Eq.(\ref{eqr}), are {\it 
 relative} corrections, we expect that the importance of 
the $W$-off-shellness effect 
is reduced.  We use the analytic computation of 
Refs.~\cite{blok1,blok2} performed as an expansion in $m_W/m_t$, 
to infer numerical values for the coefficients.
We employ the  input parameters $m_t = 172.85~{\rm GeV}$ 
and $m_W = 80.419~{\rm GeV}$  to obtain \cite{blok1,blok2}
\be
\begin{split} 
& C_F T_R \Delta_{l} = 7.978,
\;\;\; C_F T_R \Delta_h = -0.1166,\;\;\;
%\\
%& 
C_F^2 \Delta_{A} = 23.939,
\;\;\;
C_F C_A \Delta_{NA} =  - 134.24.
\end{split}
\label{eqr1}
\ee
We compare these results of the analytic computation \cite{blok1,blok2}
with our 
{\it numerical} results. 
We find\footnote{
All numerical results  presented in this Section 
are  obtained using the
adaptive Monte-Carlo algorithm VEGAS~\cite{lepage} as implemented 
in the CUBA library~\cite{hahn}. }
\be
\label{eqr2}
\begin{split} 
& C_F T_R \Delta_{l}^{\rm num} = 7.970(6),
\;\;\; C_F T_R \Delta_h^{\rm num}  = -0.11662(1),\;\;\;
\\
& 
C_F^2 \Delta_{A}^{\rm num} = 24.38(25),
\;\;\;\;\;
C_F C_A \Delta_{NA}^{\rm num} =  - 133.6(4),
\end{split}
\ee
where integration errors are explicitly shown.
By comparing Eq.(\ref{eqr1}) and Eq.(\ref{eqr2}), 
we find good agreement between results of 
analytic and 
numerical computations.  The level of agreement is rather  striking 
since there are  large cancellations 
between various contributions to the NNLO final result. To illustrate 
this point, we show individual contributions to the abelian coefficient $C_F^2 \Delta_{A}^{\rm num}$
with their integration errors.  They read 
\be
\begin{split}
& C_F^2 \Delta_{A, \rm RR} = -127.12(23),\;\;\;
\;\;\;\;C_F^2 \Delta_{A,\rm RV} = 60.24(6),
\;\;\;\;C_F^2 \Delta_{A,\rm VV} = 83.0154(5),
\nonumber \\
& C_F^2 \Delta_{A,\rm ren} = 16.513(12),
\;\;\;\;\;\;\;\;\;C_F^2 \Delta_{A,q \bar q} = -8.227(2),
\\
& C_F^2 \Delta_{A} = C_F^2 \Delta_{A, \rm RR} 
+ C_F^2 \Delta_{A,\rm RV} 
+ C_F^2 \Delta_{A,\rm VV} 
+ C_F^2 \Delta_{A,\rm ren}
+ C_F^2 \Delta_{A,q \bar q} = 24.38(25),
\end{split} 
\ee
where we include double-real (RR), real-virtual (RV), two-loop virtual (VV), 
renormalization and quark-pair corrections. 
Comparing individual contributions with the final result for $\Delta_A$, 
we observe very significant cancellations. It is therefore quite remarkable 
that  we can maintain ${\cal O}(1\%)$ error on the NNLO coefficient. 

\begin{figure*}[t]
\centering
\includegraphics[scale=0.6]{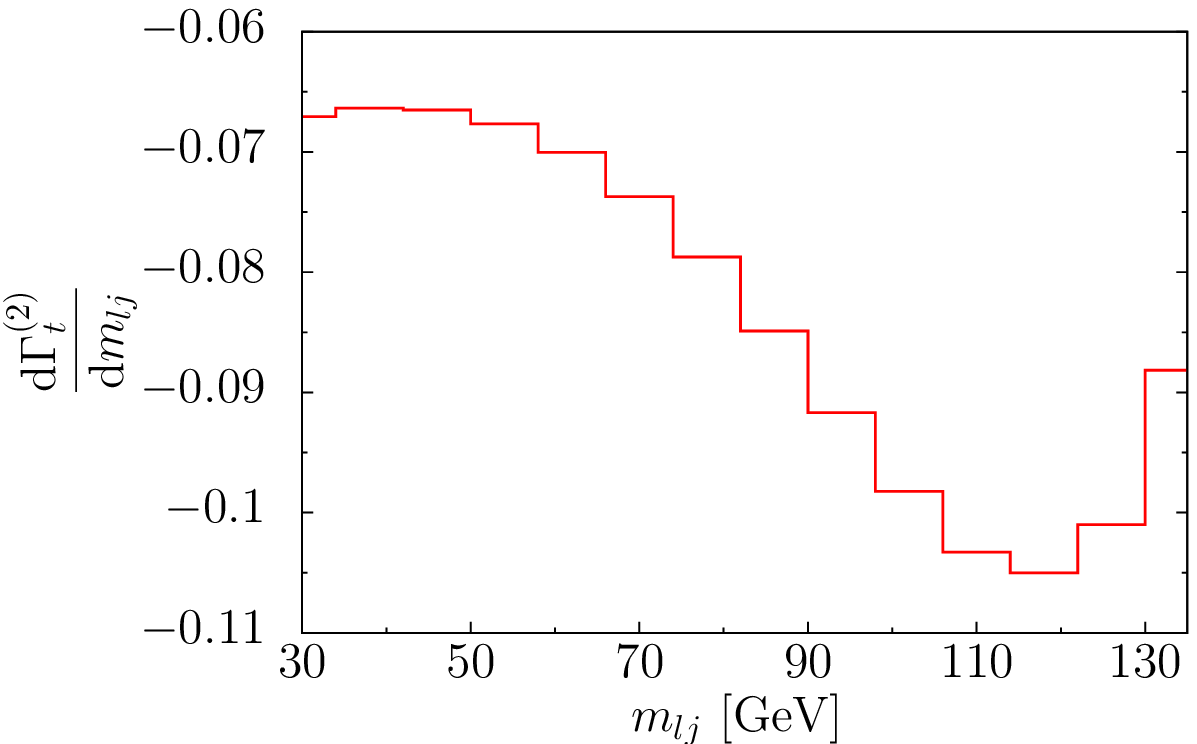}\;\;\;\;\;
\includegraphics[scale=0.6]{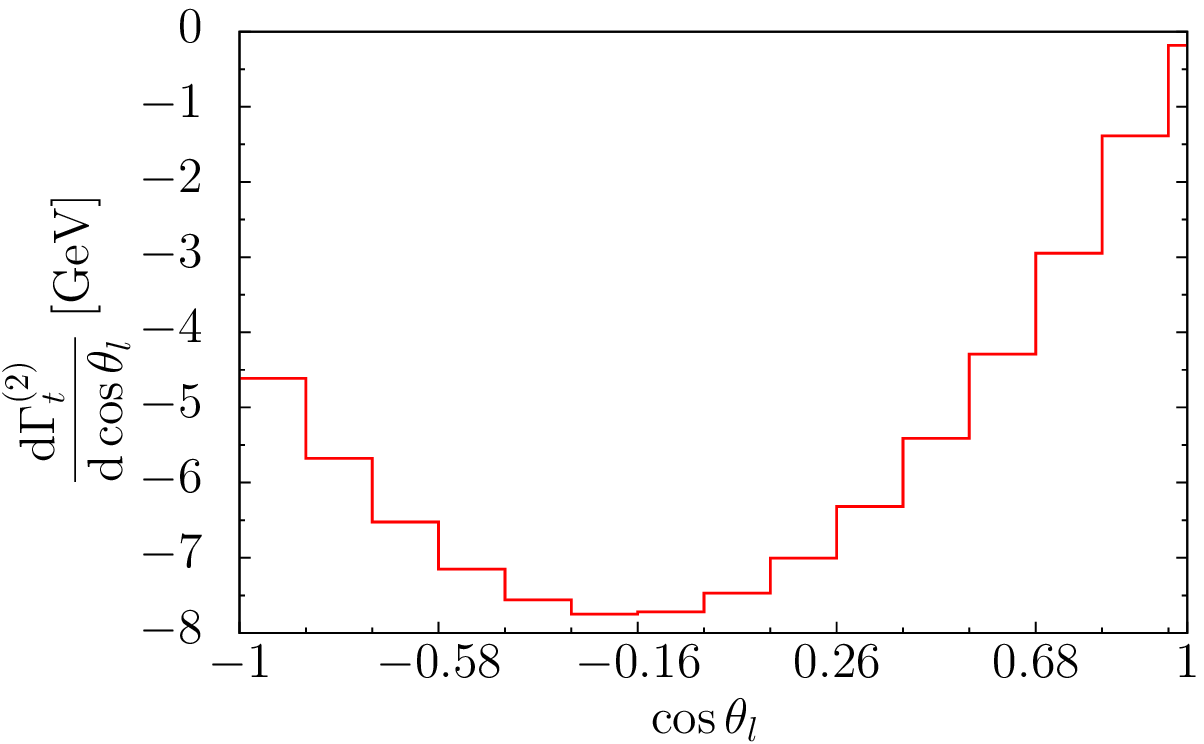}
\caption{
Left pane: distribution of the second order coefficient 
${\rm d} \Gamma_t^{(2)}$ in  
invariant mass of the positron and the hardest jet. Right pane:  
distribution of the second order coefficient 
${\rm d} \Gamma_t^{(2)}$ in the opening angle of the positron 
with respect to the $W$-direction of motion, in the $W$-rest frame. 
See text for details. 
}\label{mljangle}
\end{figure*}

We now turn to the discussion of 
 kinematic distributions. To facilitate this discussion,  
we write the differential decay rate as
\be
{\rm d} \Gamma_t = {\rm d} \Gamma_t^{(0)}
 + \left ( \frac{\alpha_s}{ 2 \pi } \right ) {\rm d} \Gamma_{t}^{(1)} 
 + \left ( \frac{\alpha_s}{2\pi} \right )^2 {\rm d} \Gamma_t^{(2)}+
{\cal O}(\alpha_s^3),
\ee
and plot $  
{\rm d} \Gamma_t^{(2)}/{\rm d} x$ 
in Figs. \ref{enrg}, \ref{mljangle} 
for various observables  $x$.  In particular, in 
Fig.~\ref{enrg} we show $
{\rm d} \Gamma_t^{(2)}/{\rm d} E_l$, 
where $E_l$ is the positron energy.  
The final result for NNLO QCD corrections arises 
as a result of large  cancellations 
between various contributions,  most notably the double-real and the 
real-virtual 
ones. To illustrate this point, we show these contributions individually in 
the left pane and the sum of all contributions to 
${\rm d} \Gamma_t^{(2)}/{\rm d} E_l$ in the right pane.  We note that 
at smaller values of $E_l$ the final result is about ten times smaller 
than double-real and real-virtual contributions separately. At this level 
of cancellations, bin-bin fluctuations that are barely seen for double-real 
and real-virtual contributions individually (left pane in Fig.~\ref{enrg})
become a serious issue.  To overcome 
it, we choose to fit individual contributions to the sum of Legendre polynomials
and then add fitted results rather than original histograms. Upon doing 
that, we find that our final result for
${\rm d} \Gamma_t^{(2)}/{\rm d} E_l$ exhibits a re-assuring 
 pattern. 
Suppose, we account for  $N_L$ Legendre polynomials in the fit. 
If $N_L$ is too small, we
find that the resulting ${\rm d} \Gamma_t^{(2)}/{\rm d} E_l$ is not stable 
against including more polynomials into the fit. If $N_L$ is too large, 
we fit bin-bin fluctuations and do not gain anything. 
However, we find that there is a range of  intermediate values of $N_L$ 
that we can use in the fit so that, on one hand, our final result for 
${\rm d} \Gamma_t^{(2)}/{\rm d} E_l$
does no depend on the exact value of $N_L$   and, on the other hand, 
 the resulting  distribution is smooth. Distributions shown in the right 
pane of Fig.~\ref{enrg} and in Fig.~\ref{mljangle} are obtained 
following this procedure.

In the left pane of Fig.~\ref{mljangle} we show NNLO QCD contributions to 
the kinematic distribution in the invariant mass of the positron and 
the hardest (in energy) jet in the event. The 
jet here is defined with the lepton 
collider $k_\perp$-algorithm where the distance between two partons 
$i$ and $j$  
is given by $y_{ij} = 2 {\rm min}(E_i^2/m_t^2,E_j^2/m_t^2)(1-\cos \theta_{ij})$.
The relative angle $\theta_{ij}$ is defined in the top quark rest frame. 
For numerical computations, we take $y_{ij} = 0.1$. 
In the right  pane of Fig.~\ref{mljangle} 
we show NNLO QCD correction  to 
the kinematic distribution of  the positron 
polar angle defined in the $W$-boson 
rest frame, relative to the direction of motion of 
the $W$-boson\footnote{The momentum of the $W$-boson 
 can be determined from the momentum of the recoiling  
hadronic system in top decay.}.
This distribution is interesting because it allows us to determine 
helicity fractions of the $W$-bosons in top decays. Indeed, to all orders 
in QCD perturbation theory, the decay rate  can be written as 
\be
\frac{{\rm d} \Gamma_t}{{\rm d} \cos \theta_l} 
= \frac{3}{4}  \sin^2 \theta_l \Gamma_L 
+ \frac{3}{8} \left ( 1+ \cos \theta_l \right )^2 \Gamma_+ 
+ \frac{3}{8}  \left (1 - \cos \theta_l \right )^2 \Gamma_-.
\ee
The  widths  $\Gamma_L,\Gamma_{\pm}$ define partial decay rates 
into polarized $W$-bosons. The helicity fractions are constructed 
from partial widths as 
$
F_{\pm,L} = \Gamma_{\pm,L}/\Gamma_t,
$
where $\Gamma_t = \Gamma_{+} + \Gamma_{-} + \Gamma_L$.
Our result  for  ${\rm d} \Gamma_t/{\rm d} \cos \theta_l$ 
shown in Fig.~\ref{mljangle} allows us to compute the NNLO QCD 
corrections to the helicity fractions.  Upon doing so, we find 
good agreement with  similar results presented in Ref.~\cite{hfrac}.
For example, by fitting the angular distribution shown in 
the right pane of Fig.~\ref{mljangle} we find the 
NNLO QCD {\rm contributions}  to 
helicity fractions\footnote{The exact definition of the helicity 
fractions and values of $\alpha_s$ used to obtain these 
results can be found in Ref.~\cite{hfrac}.} 
$\left [ \delta F_L, \delta F_-, \delta F_+ \right ] = 
[-0.0022(1),0.0021(1),0.0001(1)]$.
These numbers should be compared to the results of analytic 
computations reported in   Ref.~\cite{hfrac},
$\left [ \delta F_L, \delta F_-, \delta F_+ \right ] = 
[-0.0023,0.0021,0.0002]$. A good agreement between the two results 
is obvious.

\section{Conclusions} 
\label{conc}

In this paper we described a computation of NNLO QCD corrections 
to semileptonic decays of the top quark at a fully-differential level. 
We have used a framework described in 
Refs.~\cite{Czakon:2010td,Czakon:2011ve,Boughezal:2011jf} that combines 
sector-decomposition and Frixione-Kunszt-Signer phase-space partitioning 
to disentangle overlapping kinematic regions that may lead to soft and 
collinear singularities. We have shown that this technique works 
in the same way as parton level Monte Carlo integrators at 
next-to-leading order, so that  weighted events can be generated 
and a variety of 
kinematic distributions can be computed in a single run of the program. 

There is  a number of phenomenological applications of the calculation 
reported in this paper that we envision. Eventually, it will be combined 
with the perturbative QCD description of the top quark production process, 
to provide NNLO  QCD results for $pp \to t \bar t \to W^+W^-b \bar b$ 
in the narrow  width approximation.  It can also be used to investigate 
the dependence of 
various observables studied in top quark decays, for example the helicity 
fractions, on  kinematic cuts applied to top quark decay products. 
A minor modification of our calculation turns it into 
a NNLO QCD result for the most important contribution to the single 
top production process in hadron collisions. 
Finally, as explained in the Introduction, our calculation 
 can be adapted to 
explore the impact of NNLO QCD corrections to inclusive charmless semileptonic 
decays of $b$-quarks  on  
the determination of the CKM matrix element $|V_{ub}|$. 
We plan to return to these applications in the near future. 

\vspace*{0.5cm}
{\bf Acknowledgments}
We are grateful to T.~Hahn for advises on CUBA 
and to F.~Tramontano for answering questions 
about Ref.~\cite{Campbell:2005bb}. 
We would like to thank J.~Gao, C.~S.~Li and H.~X.~Zhu for pointing out an error in one of the plots
of the original version of this manuscript. 
The research of F.C. and K.M.  is supported by US NSF under grant PHY-1214000.
The  research of K.M. and M.B. 
is partially supported by Karlsruhe Institute of Technology through a grant provided  
by its Distinguished Researcher Fellowship Program. 
The research of M.B. is partially supported by the DFG through the SFB/TR~9 ``Computational
Particle Physics''.
Calculations described in this paper were performed at the Homewood 
High Performance Computer Cluster at Johns Hopkins University. 

\appendix

\section{Tree-level amplitudes}
\label{tree}

In this Appendix we collect formulas 
for  helicity amplitudes for top quark decay processes
$t \to b + n_g g + e^+ + \nu$, for  $n_g \le 2$. 
We note that we prefer to use helicity amplitudes 
for reasons of  speed and efficiency of the 
computation. While for top quark decay  this is not a critical issue -- the 
number of diagrams is sufficiently small even at NNLO -- it becomes essential 
if one considers applicability of similar techniques to more complicated 
processes.    Some amplitudes that we present below 
have already  appeared in the literature 
\cite{Campbell:2012uf} while others are new. 

To deal with the massive top quark, we follow the  standard procedure
that is based on the decomposition of a time-like four-momentum into a sum of 
two light-like four-momenta \cite{kleiss}. To accomplish this, we choose 
an arbitrary light-like momentum $\eta$ and write 
\be
p_t = p_1 + \frac{m_t^2}{s_{p_t\eta}} \eta,
\ee
where $ s_{ij} =  2 p_i\cdot p_j$.
We note  that $p_1^2=0$ 
by construction.  We can now write the top quark spinor as
\be
\begin{split} 
& u_-(p_t) = (\not p_t + m_t)|\eta\rangle \frac{1}{\spaa {1} \eta} 
= |1] + |\eta\rangle \frac{m_t}{\spaa {1} \eta},
\\
& u_+(p_t) = (\not p_t + m_t)|\eta] \frac{1}{\spbb {1} \eta} 
= |1 \rangle + |\eta] \frac{m_t}{\spbb {1} \eta},
\end{split} 
\ee
where we use the standard notation for spinor-helicity variables. 
In particular, for $p_i^2 = 0$, 
$|i \rangle = |p_i \rangle = u_+(p_i)$ and  
$|i ] = | p_i ] = u_-(p_i)$. 
It is convenient to choose the auxiliary momentum $\eta$ to 
be the positron momentum $p_6$ to compute helicity amplitudes. 

We now present results for helicity amplitudes starting 
from the tree-level process 
$t(p_t)\to b(p_2)+\nu(p_5)+e^+(p_6)$.
We introduce the Breit-Wigner denominator $D_W(p_W^2) = 
(p_W^2 - m_W^2 + i m_W \Gamma_W)^{-1}$ and 
write the amplitude as 
\be
\mathcal A_{h_t}(p_t,p_2;p_5,p_6) = g_W^2 D_W(s_{56}) 
\delta_{i_1 i_2} A_{h_t}(1,2;5,6) \equiv \mathcal F \delta_{i_1 i_2} 
A_{h_t}(1,2;5,6).
\ee
The amplitude $A_{h_t}(1,2;5,6)$ is the color-stripped amplitude and it only depends 
on the helicity of the top quark $h_t$ since helicities of all massless fermions 
are fixed thanks to the left-handed nature of the weak current. We find 
\be
A_+(1,2;5,6) = 0,
\;\;\;A_-(1,2;5,6) = \spaa 2 5 \spbb 6 1 .
\ee
In this case the sum over colors and helicities is trivial; the amplitude 
squared reads
\be
\sum_{\rm col,hel}\mathcal A_{h_t}(p_t,p_2;p_5,p_6) = \left| \mathcal F\right|^2 N_c \left| A_-(1,2;5,6)\right|^2.
\ee

Next, we consider amplitudes for the process 
$t(p_t)\to b(p_2)+g(p_3)+\nu(p_5)+e^+(p_6)$.
We write it as 
\be
\mathcal A_{h_t h_g}(p_t,p_2,p_3;p_5,p_6) 
= g_s {\mathcal F} T^{a_3}_{i_2 i_1} A_{h_t h_g}(1,2,3;5,6),
\ee
where  $T^a$ is an $SU(3)$ color matrix  normalized as $\Tr(T^a T^b)=\delta^{ab}$ and $h_{t,g}$ are helicities of the top quark and of the gluon, respectively. 
We obtain the following results for the color-stripped 
helicity amplitudes
\be
\begin{split} 
& A_{--}(1,2,3;5,6) =  
-\frac{[6 1] \langle 3 |\hat p_{56}| 2] \langle 2 5\rangle }{\spab 3 {|\hat p_t|} 3
   [3 2]}-\frac{[6 1] \langle 3 5\rangle }{[32]}, \\
& A_{-+}(1,2,3;5,6) =  -\frac{[61] \langle 2|\hat p_{56}|3] 
\langle 2 5\rangle }{\langle
   3|\hat p_t|3] \langle 2 3\rangle }-\frac{[3 1] [6 3] \langle
   2|5\rangle }{\langle 3|\hat p_t|3]}, \\
& A_{++}(1,2,3;5,6) =  
-m_t  \frac{\spaa 2 5 {\spbb 6 3 }^2}{\spbb 6 1 \spab 3{|\hat p_t|} 3},\\
& A_{+-} = 0,
\end{split}
\ee
where $p_{56} = p_5 + p_6$.
The sum over colors is trivial also in this case. We find
\be
\sum_{ \rm col} |\mathcal A|^2 = |\mathcal F|^2 g_s^2 (N_c^2-1) |A|^2 = 2 N_c C_F |\mathcal F|^2 g_s^2 |A|^2 .
\ee

The last set of tree amplitudes that we require for our calculation 
are the ones that describe the process with two gluons in the 
final state $t(p_t)\to b(p_2)+g(p_3)+g(p_4)+\nu(p_5)+e^+(p_6)$.
We write 
\be
\mathcal A_{h_t h_{g_3} h_{g_4}}(p_t,p_2,p_3,p_4;p_5,p_6) = \mathcal F g_s^2 \sum_{\sigma}\lp T^{a_{\sigma_3}}\cdot T^{a_{\sigma_4}}\rp _{i_2 i_1} 
A_{h_t h_{g_{\sigma_3}} h_{g_{\sigma_4}}}(1,2,\sigma(3),\sigma(4);5,6),
\ee
where $\sigma$ denotes two permutations of the gluons $g_3,g_4$.
The color-stripped helicity amplitudes are computed 
to be
\be
\begin{split}
& A_{---} = \frac{[61] \langle 25\rangle 
   [2|\hat p_t \hat p_{34}|2]}{\tilde s_{t34} [32] [42]
   [43]}-\frac{[61] \langle 4|\hat p_{56}|2] \langle 25\rangle 
   \langle 3|\hat p_t|2]}{\tilde s_{t34} [32] [42]
   [3|\hat p_t|3\rangle } 
\\
  & 
+\frac{[61] \langle 45\rangle 
   \langle 3|\hat p_t|2]}{[32] [42] [3|\hat p_t|3\rangle
   }-\frac{[61] \langle 45\rangle }{[32] [43]}-\frac{[61]
   \langle 35\rangle }{[42] [43]}, \\
& A_{--+} = \langle 25\rangle  
\left(\frac{[61]}{[43] \langle 34\rangle }{ \left(-\frac{[42] \langle
   23\rangle ^2}{s_{234} \langle 2|4\rangle
   }+\frac{\langle 3|\hat p_t|4]^2}{\tilde s_{t34} \langle
   3|\hat p_t|3]}+\frac{\langle 2|3\rangle  \langle
   3|\hat p_t|4]}{\langle 3|\hat p_t|3] \langle 2|4\rangle
   }\right)} \right. \\
       & \left. + \frac{[4|1] [6|4]
   \langle 3|\hat p_t|4]}{\tilde s_{t34} [4|3] \langle
   3|\hat p_t|3]}\right)-\frac{[6|1] \langle 2|3\rangle ^2 \langle
   3|5\rangle }{s_{234} \langle 2|4\rangle  \langle
   3|4\rangle }, \\
& 
A_{-+-} = \frac{[61] \langle 25\rangle}{[43] \langle 34\rangle }
\left(\frac{[32] \langle 4|\hat p_t|3]}{[42] \langle 3|\hat p_t|3]}
-\frac{[32]^2 \langle 24\rangle }{s_{234} [42]}+\frac{\langle 4|\hat p_t|3]^2}{{\tilde s_{t34}} \langle 3|\hat p_t|3]}\right) + \nn\\
&+\frac{[61] \langle 45\rangle }{[42] \langle 34\rangle }\left(\frac{[32] \langle 24\rangle }{s_{234}} -\frac{\langle
   4|\hat p_t|3]}{\langle 3|\hat p_t|3]}\right) 
 +\frac{[63] [31] \langle 25\rangle }{[43] \langle 3|\hat p_t|3]}\left(-\frac{\langle 4|\hat p_t|3]}{{\tilde s_{t34}}}-\frac{[32]
   }{[42]}\right) +\frac{[31] [63] \langle 45\rangle }{[42] \langle 3|\hat p_t|3]},
 \\
& A_{-++} = \frac{\langle 25\rangle}{\tilde s_{t34}}
  \left(\frac{[61] \langle 2|\hat p_{56}|4] \langle 4|\hat p_t|3]}{\langle 24\rangle  \langle 34\rangle 
   \langle 3|\hat p_t|3]}-\frac{[31] [63] \langle 2|\hat p_{56}|4]}{\langle 24\rangle  \langle 3|\hat p_t|3]}+\frac{[61] \langle
   2|\hat p_t|3]}{\langle 24\rangle  \langle 34\rangle }+ \right. \\ 
& +\left.\frac{[41] [64] \langle 4|\hat p_t|3]}{\langle 34\rangle  \langle
   3|\hat p_t|3]}-\frac{[31] [43] [64]}{\langle 3|\hat p_t|3]}+\frac{[31] [63] 
\langle 23\rangle }{\langle 24\rangle  \langle
   34\rangle }+\frac{2 [31] [64]}{\langle 34\rangle }\right),
\end{split} 
\ee
and 
\be
\begin{split} 
& A_{+--} = 0, \\
& A_{+-+} = m_t\frac{[64]^2 \langle 25\rangle  
\langle 3|\hat p_t|4]}{\tilde s_{t34} [43] [61] \langle 3|\hat p_t|3]}, \\
& A_{++-} = m_t\frac{[63]^2}{[61] \langle 3|\hat p_t|3]}\lp
-\frac{\langle 25\rangle  
\langle 4|\hat p_t|3]}{\tilde s_{t34} [43]}
+\frac{\langle 45\rangle }{[42]}-\frac{[32] \langle 25\rangle
   }{[42] [43]}\rp , \\
& A_{+++}= 
m_t\frac{\langle 25\rangle }{\langle 34\rangle  \langle 3|\hat p_t|3]}
\lp
\frac{[43] \langle 1|\hat p_{34}|6]}{\tilde s_{t34}}+\frac{[63] \langle 2|\hat p_{34}|6]}{[61] \langle 24\rangle }
\rp,
\end{split}
\ee
with $\tilde s_{t34}= s_{34} - s_{3t}-s_{4t}$. The color-summed amplitude 
now reads
\bea
\sum_{ \rm col} |\mathcal A|^2 
& = & 4 N_c |\mathcal F|^2 g_s^4 C_F \Bigg [C_F \lp |A(1,2,3,4;5,6)|^2 + |A(1,2,4,3;5,6)|^2 \rp  +  \nn\\
& + &  \lp C_F -\frac{C_A}{2}   \rp 2{\rm Re} \left[ A(1,2,3,4;5,6) A^*(1,2,4,3;5,6) \right]  \Bigg ].
\eea

\section{Calculation of one-loop  integrals for the soft limit}
\label{oneloopMI}
In this Appendix, we give an example of how one-loop eikonal 
integrals needed for the soft current at one-loop can be computed. 
Consider the four-point function, defined in Eq.(\ref{masterintegrals})
\be
I_4 = \int \frac{d^d k}{i (2 \pi)^d} \frac{1}{(k-p_3)^2 k^2 (2 p_b \cdot k)
 (2 p_t \cdot k - 2 p_t  \cdot p_3)},
\ee
where $p_t^2 = m_t^2$, $p_b^2 = p_3^2 = 0$. 
We assume that each propagator contains a term $+i0$ which defines 
the proper analytic continuation in cases of branch cuts. 

To compute $I_4$, we first combine $k^2$ and $(k-p_3)^2$ using Feynman 
parameters 
\be
\frac{1}{k^2 (k-p_3)^2} = \int \limits_{0}^{1} 
\frac{{\rm d} x }{[(k-x p_3)^2]^2}.
\label{aeq1}
\ee
Next, we exponentiate the eikonal quark propagators introducing the integrals 
over ``proper time''
\be
\frac{1}{(2p_t \cdot k - 2 p_t \cdot p_3 + i0)(2 k \cdot p_b + i0)}
= - \int \limits_{0}^{\infty} 
{\rm d} \tau_1 {\rm d} \tau_2  e^{i 2 k \cdot p_b \tau_1 
+ i (2 p_t \cdot k - 2 p_t \cdot p_3) \tau_2}.
\label{aeq2}
\ee
We now combine Eqs.(\ref{aeq1}, \ref{aeq2}) and shift the loop momentum 
$k \to k + p_3 x$. The resulting integral becomes 
\be
I_4 = 
i \int \limits_{0}^{\infty} 
{\rm d} \tau_1 {\rm d} \tau_2 
\int \limits_{0}^{1} {\rm d} x 
e^{i ( -2 p_t \cdot p_3 (1-x) \tau_2 + 2 p_3 \cdot p_b x \tau_1 )}
\int \frac{{\rm d}^d k}{(2\pi)^d} \frac{1}{[k^2]^2}
e^{i k (2 p_t \tau_2 + 2 p_b \tau_1)}.
\ee
The integral over $k$ is easily evaluated
\be
\int \frac{{\rm d}^d k}{(2\pi)^d} \frac{1}{[k^2]^2}
e^{i k (2 p_t \tau_2 + 2 p_b \tau_1)} 
= \frac{i^{1+2\ep} \Gamma(-\ep)}{(4\pi)^{d/2}} 
[ m_t^2 \tau_2^2 + 2 p_t \cdot p_b \tau_2 \tau_1 ]^{\ep}.
\ee
To proceed  further, we change variables $\tau_1  = \tau_2 \xi$, 
$ 0 \le \xi \le \infty$ and integrate over $\tau_2$ and $x$.   
We obtain 
\be
\begin{split}
I_4 = & -\frac{\Gamma(-\ep)  \Gamma(1+2\ep)}{(4\pi)^{d/2}}
\int \limits_{0}^{\infty} 
\frac{{\rm d} \xi \; (m_t^2 + 2 p_t \cdot p_b \xi)^{\ep}}{ (2 p_t \cdot p_3 
+ 2 p_b \cdot p_3 \xi)}
 \left( \frac{e^{2 \pi i \ep}}{(2 p_b \cdot p_3 \xi)^{1+2\ep}} 
+ \frac{1}{(2 p_t \cdot p_3)^{1+2\ep}} \right).
\end{split} 
\ee
The final integration over $\xi$ is easy to perform in terms of hypergeometric 
functions. The result is shown in Eq.(\ref{eq_me}).

\end{document}